\documentclass[aip,  jmp,
amsmath,amssymb,
preprint,%
]{revtex4-1}

\usepackage{graphicx}
\usepackage{dcolumn}
\usepackage{bm}

\usepackage[utf8]{inputenc}
\usepackage[T1]{fontenc}
\usepackage{mathptmx}
\usepackage{etoolbox}

\makeatletter
\def\@email#1#2{%
 \endgroup
 \patchcmd{\titleblock@produce}
  {\frontmatter@RRAPformat}
  {\frontmatter@RRAPformat{\produce@RRAP{*#1\href{mailto:#2}{#2}}}\frontmatter@RRAPformat}
  {}{}
}%
\makeatother

\usepackage{mathrsfs}
\usepackage{color}

\usepackage[normalem]{ulem}

\newtheorem{theorem}{{\bf Theorem}}

\newtheorem{lemma}{{\bf Lemma}}
\newtheorem{remark}{{\bf Remark}}
\newtheorem{example}{{\bf Example}}

\def \rmd{\mathrm{d}}

\def \ad{\mathrm{ad}}
\def \Ad{\mathrm{Ad}}
\def \vol{\mathrm{vol}}

\def\bell{\boldsymbol{\ell}}
\def \bmu{\bm{\mu}}
\def \bomega{\bm{\omega}}
\def \bOmega{\bm{\Omega}}
\def \bzeta{\bm{\zeta}}

\newcommand{\bi}[1]{\bm{#1}}

\begin{document}
\title{The kinetic origin of the fluid  helicity 
-- a symmetry in the kinetic phase space 
}

\author{Zensho Yoshida}
 \altaffiliation[Also at ]{National Institute for Fusion Science, Toki, Gifu 509-5292, Japan}
 \affiliation{Graduate School of Frontier Sciences, University of Tokyo, Kashiwa, Chiba 277-8561, Japan}%

\author{Philip J. Morrison}%
\affiliation{Department of Physics and Institute for Fusion Studies, University of Texas at Austin, TX 78712-1060, USA}%

 \email{yoshida@ppl.k.u-tokyo.ac.jp, ~morrison@physics.utexas.edu}

\date{\today}

      



\begin{abstract} 
Helicity,  a  topological  degree that measures  the winding and linking of vortex lines,  is preserved by  ideal (barotropic) fluid dynamics.   In the context of the Hamiltonian description, the helicity is a Casimir invariant characterizing a foliation of the associated Poisson manifold.   Casimir invariants are special invariants that depend on the Poisson bracket, not on  the particular choice of the Hamiltonian.  The  total mass (or particle number) is another Casimir invariant, whose invariance guarantees the mass (particle) conservation (independent of any specific choice of the Hamiltonian). In a kinetic description (e.g.\  that of the Vlasov equation),  the helicity is no longer an invariant (although the total mass remains a Casimir of the Vlasov's Poisson algebra).  The implication is that some ``kinetic effect'' can violate the constancy of the helicity.
To elucidate how the helicity constraint emerges or submerges,
we examine the fluid reduction of the Vlasov system;
the fluid (macroscopic) system is a ``sub-algebra'' of the kinetic (microscopic) Vlasov system.
In the Vlasov system, the helicity can be conserved, if a special \emph{helicity symmetry} condition holds.
To put it another way, breaking  helicity symmetry induces a change in the helicity.
We delineate the geometrical meaning of  helicity symmetry,
and show that, for a special class of flows (so-called epi-2 dimensional flows), 
the helicity symmetry is written as $\partial_\gamma =0$ for a coordinate $\gamma$ of the configuration space.


\end{abstract}



\maketitle

%


\section{Introduction}
\label{sec:introduction}

The notion of helicity appears in both ideal fluid mechanics and ideal magnetohydrodynamics (MHD).  In  fluid mechanics,  ideas pertaining to helicity and its conservation date back to Lord Kelvin in the nineteen century, while its recognition as an indicator of topological linkage of vortex lines was given in Ref.\;\onlinecite{Moreau}.   Similarly, that a kind  of  helicity is conserved by the ideal MHD equations was noted in  Ref.\;\onlinecite{Woltjer},  but its  topological recognition in terms of magnetic field line linkage was given in  Ref.\;\onlinecite{Moffatt1969}. The study of helicity in a variety of fluid  and magnetofluid contexts continues to fascinate researchers. For example, recently the  role helicity plays in the reconnection of vortex tubes has been experimentally observed  in real  fluids (e.g.\  Refs.\;\onlinecite{Irvine2017,Irvine2018})  and the role it plays in cascades of  turbulence has  been investigated in a variety of  numerical simulations (e.g.\  Refs.\;\onlinecite{Biferale2017,Biferale2020}).   However, to our knowledge there have been no studies of how helicity emerges from  the kinetic description that underlies the fluid description, which is the subject matter of the present paper. 

Both ideal fluid mechanics and collisionless kinetic theory possess noncanonical Hamiltonian structure,\cite{Morrison1998} with Poisson brackets that generate flows on Poisson manifolds.\cite{Weinstein1983}  Consequently, both fluids and kinetic theories possess  Casimir invariants,  universal invariants independent of any particular choice of the Hamiltonian for the respective theories.  Therefore these invariants
represent  types of \emph{topological constraints} inherent to  Poisson manifold   phase spaces.
Every orbit is constrained to Casimir leaves,  level-sets  of Casimirs,
so the gradient of a Casimir is transverse to the leaf its constancy defines. 
This means that the gradient of the Casimir belongs to the kernel of the Poisson matrix (the 2-vector that maps the gradient of the Hamiltonian to the Hamiltonian vector field) that defines the Poisson bracket.
By ``degenerate''  we mean that the Poisson matrix defined on the Poisson manifold  has a nontrivial kernel.
We note that the element of the kernel (covector) is not necessarily \emph{integrable}, i.e.,  the gradient of some scalar.   However, a Casimir is such an integral, yielding  a foliation of the Poisson manifold by its level-sets.

When a Casimir is given,  one can interpret it as an \emph{adiabatic invariant} made ``variable'' by adding  an \emph{angle variable} to complete a conjugate action-angle pair.\cite{FDR2014}
After embedding the noncanonical Poisson manifold into the inflated phase space,
the constancy of the Casimir can then be  re-interpreted as arising from a  symmetry with respect to the supplementary angle variable; hence, its constancy is now attributed to this  specific symmetry of the Hamiltonian.
The  example of Sec.\,\ref{sec:so(3)} will delineate such a relation from the opposite viewpoint: starting from the canonical (symplectic) Poisson manifold $\mathfrak{sp}(6,\mathbb{R})$,
we derive the noncanonical $\mathfrak{so}(3)$ Lie-Poisson algebra by \emph{reduction}.\cite{Marsden}  
Restricting  the 6 dimensions of $\mathfrak{sp}(6,\mathbb{R})$, represented by the position vector $\bi{q}$ and the momentum vector $\bi{p}$, to 3 dimensions  determined  by the angular momentum $\bell=\bi{q}\times\bi{p}$,
the system reduces to the 3-dimensional $\mathfrak{so}(3)$ Lie-Poisson manifold with the magnitude $|\bell|$ becoming  the Casimir of the $\mathfrak{so}(3)$ Lie-Poisson bracket.  Consequently,   the effective available phase space shrinks to the 2 dimensional spherical surface $|\bell|=$ constant.
A physical example of such a reduced angular momentum system is the \emph{Euler top}, which is a point mass bound to the origin of the coordinate space by a rigid, mass-less rod.
Then, the angle between the position $\bi{q}$ and the momentum $\bi{p}$ is fixed to be perpendicular.
If this angle is allowed to vary (for example, if the rod is not sufficiently rigid), 
the constancy of  the Casimir $|\bell|$ is broken (see Sec.\,\ref{sec:so(3)}),
i.e.,    ``rigidity'' is the root cause of the Casimir.
More precisely, for the Casimir to be invariant, there must be a distinct separation of the time scale (or energy) between the dynamics of the top and the change of the angle variable.
Consequently, from this point of view we may interpret the Casimir as an adiabatic invariant.

In fluid mechanics, the \emph{helicity} is a Casimir of the Hamiltonian formalism of the ideal (barotropic) fluid model,\cite{Morrison-Greene1980} which is a  measure of  the winding and linking of vortex lines.\cite{Moreau,Moffatt1969}
Interestingly, in a kinetic description (e.g.\  the Vlasov equation; see Sec.\,\ref{sec:Vlasov}), 
the helicity is no longer an invariant.
This implies that some ``kinetic effect'' can violate the topological constraint associated with  the helicity.
It is known that the ideal fluid system can be formulated as a reduction (subalgebra) of the Vlasov system.\cite{Morrison1998}
In the Vlasov system,  the helicity can still be conserved, if a special \emph{helicity symmetry} condition holds. To put it  another way, braking of the helicity symmetry allows for a changing  helicity.
The aim of this work is to elucidate the geometrical meaning of the helicity symmetry,
and study how the topological constraint associated with the helicity invariance can be broken in kinetic theory.
For a special class of flows (so-called epi-2 dimensional flows\cite{epi-2D}) we will show that
the helicity symmetry is written as $\partial_\gamma =0$ with $\gamma$ being a configuration space coordinate.


\section{Casimir and gauge symmetry in a ``reduced system'' -- examples}
\label{sec:examples}

Because Casimirs play a central role in this work, we explain, by simple examples, how a Casimir is  ``created'' by a \emph{reduction} of some kind, and how it is related to the \emph{gauge symmetry} of the reduction.


\subsection{Reduction of canonical variables}
\label{subsec:reduction}

We start with the canonical Hamiltonian system of a point mass moving in $\mathbb{R}^n$ with the  phase space being $M=\mathbb{R}^{2n}$.
The  coordinates of a  point of $M$ represent a  state vector,    $\bi{z}=(\bi{q},\bi{p})^{\mathrm{T}}$,  with
  position $\bi{q}$ and momentum $\bi{p}$.
On $C^\infty(M)$, the space of observables, we define the canonical Poisson bracket
\begin{equation}
[G, H ]  = \sum_{j=1}^n (\partial_{q^j} G) \, (\partial_{p_j} H) 
- (\partial_{q^j} H) \,  (\partial_{p_j} G) .
\label{sp(2n,R)}
\end{equation}
Denoting by $\partial_{ \bi{z}} G \in T^* M$ the gradient of $G \in C^\infty(M)$, and by
$\langle \bi{x} , \bi{y} \rangle$ the natural pairing of $\bi{x}\in T^*M$ and $\bi{y}\in TM$,
we may rewrite (\ref{sp(2n,R)}) as
\begin{equation}
[G, H ]  =  \langle \partial_{ \bi{z}} G, J  \partial_{ \bi{z}} H \rangle,
\label{sp(2n,R)-2}
\end{equation}
with the Poisson operator (matrix)
\begin{equation}
J = \left( \begin{array}{cc}
0 & I \\
-I& 0   \end{array} \right) \quad 
\in \mathrm{Hom}(T^*M, TM).
\label{sp(2n,R)-3}
\end{equation}
The Hamiltonian vector $\dot{\bi{z}} \in TM$ is given by 
\[
 \dot{\bi{z}} = J \partial_{\bi{z}} H.
\]


We assume $n=2$ and denote the corresponding symplectic  manifold by $M_4$ ($= \mathbb{R}^4$).
As a trivial example of reduction, we suppose all observables are independent to $q^2$ and $p_2$.
Then, the Poisson bracket evaluates as
\begin{equation}
[G, H ]_{M_2}  = 
(\partial_{q^1} G) \, (\partial_{p_1} H) - (\partial_{q^1} H) \,  (\partial_{p_1} G) \,,
\label{sp(2,R)}
\end{equation}
which defines a canonical Poisson algebra on the submanifold $M_2 = \{ \bi{z}_2=(q^1,p_1)^{\mathrm{T}} \} = \mathbb{R}^2$, which is embedded in $M_4$ as a leaf $\{\bi{z}\in M_4;\, q^2=c,\, p_2=c' \}$, where $c$ and $c'$ are arbitrary constants.

An interesting situation occurs when we only suppress the coordinate $q^2$ in the set of observables: the 
 reduced phase space is the 3-dimensional submanifold $M_3 = \{ \bi{z}_3=(q^1,p_1,p_2)^{\mathrm{T}} \}$.
For $G$ and $H$ satisfying $\partial_{q^2} G = \partial_{q^2} H =0$,   the Poisson bracket evaluates the same as (\ref{sp(2,R)}) and we may write
\[
[ G, H] _{M_3} = \langle \partial_{ \bi{z}_3} G, J  \partial_{ \bi{z}_3} H \rangle
\]
with the Poisson operator (matrix)
\[
J = \left( \begin{array}{ccc}
\;0\; & \;1\; & \;0\; \\
-1& 0 & 0 \\
0 & 0 & 0 \end{array} \right),
\]
whose rank is two.  
Therefore $M_3$ is a degenerate Poisson manifold.
The kernel of this $J$ includes the vector $(0,0,1)^{\mathrm{T}}$, which can be integrated to define
a Casimir $C= p_2$.
Therefore, the effective dimension is further reduced down to two;
the state vector $\bi{z}$ can only move on the 2-dimensional leaf $M_2$.
Evidently the ``freezing'' of $C=p_2$ is due to the suppression of its conjugate variable $q^2$.

When we observe $M_3$ from $M_4$, the reduction (i.e. the suppression of the coordinate $q^2$ in the observables) means the \emph{symmetry} $\partial_{q^2}=0$.
As usual, the integral of motion $p_2$ in $M_4$ arises because ${q^2}$ is ignorable, i.e., if the Hamiltonian has the symmetry $\partial_{q^2} H=0$.

The variable $q^2$ conjugate  to the Casimir $C=p_2$ can be regarded as a \emph{gauge parameter}.
The gauge group (denoted by $\Ad_C$), which does not change the submanifold $M_3$ embedded in $M_4$, is
 generated by the adjoint action
\[
\ad_C = [\circ, C] = \partial_{q^2} ,
\]
implying that the \emph{gauge symmetry} is written as $\partial_{q^2}=0$.
This is evident, because the state vector $\bi{z}_3=(q^1,p_1,p_2)^{\mathrm{T}} \in M_3$ is independent of $q^2$.
Notice that $\partial_{q^2} =0$ is the symmetry producing the integral $C=p_2$ (which we may call \emph{Casimir symmetry}),
and, at the same time, the \emph{gauge symmetry} of the submanifold $M_3$.

A similar reduction occurs when we consider the canonical pair, 
\[
\mu = \frac{1}{2}\left[ (q^2)^2+ (p_2)^2 \right], \quad \theta = \tan^{-1} \left( \frac{q^2}{p_2}\right).
\]
If we suppress $\theta$ in the set of  observables, $\mu$ becomes the Casimir of $M_3'=M_4/\{\theta\}$.
The motion of a magnetized particle is an example, where $\mu$ corresponds to the magnetic moment, and $\theta$ to the gyration angle.
When the gyro period is negligibly shorter than the time scale of interest, $\mu$ can be dealt with as an adiabatic invariant. Such ``coarse graining'' means that we consider the average over $\theta\in [0,2\pi)$ and put $\ad_\mu=\partial_\theta=0$ for all observables.  
See Refs.\;\onlinecite{Northrop,Henrard} for in-depth  treatments  of magnetized charged particle dynamics and adiabatic invariance.

In the Sec.\;\ref{sec:so(3)} we consider another example  that displays a less  trivial relation between the Casimir and gauge symmetry.


\subsection{Reduction of $\mathfrak{sp}(6,\mathbb{R})$ to the $\mathfrak{so}(3)$ Lie-Poisson manifold}
\label{sec:so(3)}

In this next example we examine the reduction that produces the $\mathfrak{so}(3)$ Lie-Poisson system,
and  how its Casimir is related to the \emph{gauge symmetry},
i.e.,  the invariance of the reduced variables with respect to the transformation (gauge group action) among the original variables.

We start with the canonical Hamiltonian system of $n=3$ with Poisson bracket given by  (\ref{sp(2n,R)}).
We let $\bi{z}=(\bi{q},\bi{p})^{\mathrm{T}} \in M_6 =\mathbb{R}^6$, and  consider the system where the observables are functions of only the angular momentum: 
\begin{equation}
\bell  = \bi{q}\times\bi{p} .
\label{6D-omega}
\end{equation}
The Euler top is such an example, where the  Hamiltonian is $H(\bell)= \sum_j \ell_j^2/(2 I_j)$
with $I_1,\, I_2,\, I_3$  being the three moments of inertia.
For such a system, the effective phase space is reduced to $M_{\bell} \cong \mathbb{R}^3$.
Let us evaluate $[\;,\;]$ for observables $\in C^\infty(M_{\bell})$.
The gradient of a functional $F\in C^\infty(M_{\bell})$ is given by
\[
\delta F = \langle \partial_{\bi{q}} F, \delta {\bi{q}} \rangle +
 \langle \partial_{\bi{p}} F, \delta {\bi{p}}  \rangle
=  \langle \partial_{\bell} F, \delta {\bell} \rangle\, .
\]
Inserting $ \delta {\bell} = (\delta \bi{q})\times\bi{p} + \bi{q}\times(\delta\bi{p})$, we find
\[
\partial_{\bi{q}} F = \bi{p}\times\partial_{\bell} F
\quad
\mbox{and}
\quad 
\partial_{\bi{p}} F = -\bi{q}\times\partial_{\bell} F \,. 
\]
Therefore,
\[
[G,H ]= 
 \langle \partial_{\bell} G, \partial_{\bell} H \times \bell  \rangle =:\{G,H\}\,,
\]
which is  a Lie-Poisson bracket (see Remark\,\ref{remark:Lie-Poisson}) as follows:
\[
\{G,H \} = \langle \partial_{\bell} G, J(\bell) \partial_{\bell} H \rangle,
\]
with the Poisson operator (matrix)
\begin{equation}
J(\bell) :=  - \bell\times \circ
= \left( \begin{array}{ccc}
0 & \ell_3 & -\ell_2 \\
-\ell_3 & 0 & \ell_1 \\
\ell_2 & -\ell_1 & 0 \end{array} \right) .
\label{so(3)_J}
\end{equation}
Notice that this Poisson operator is a linear function of $\bell$, 
  the signature of a Lie-Poisson algebra (see Remark\,\ref{remark:Lie-Poisson}).
Here  $\mathrm{rank}\,J(\bell) =2$ (avoiding the point $\bell=0$ where $\mathrm{rank}\,J(\bell) =0$) so we expect a single Casimir  of the reduced Poisson algebra,  which evidently is 
\[
C =  \frac{1}{2}|\bell|^2\,,
\]
a function easily  seen to satisfy   $\{ G, C\} =0$ ($\forall G \in C^\infty(V_{\bell})$),
or $J(\bell) \partial_{\bell} C =0$.

When we take $C$ as the Hamiltonian, the adjoint action
\begin{eqnarray}
\ad_C  =[\circ , C] &=& \left(\sum_{j=1}^3 \partial_{p_j} C \partial_{q^j} - \partial_{q^j} C \partial_{p_j}\right)
\nonumber \\
&=& \bell\times\bi{q} \cdot\partial_{\bi{q}}+ \bell\times\bi{p}\cdot\partial_{\bi{p}}
\label{gauge_symmetry_of_omega}
\end{eqnarray}
generates the \emph{gauge transformation} of the reduced variable $\bell$;
by direct calculation it follows easily that $[\ell_j, C]=0$ ($j=1,2,3$).

This gauge transformation has the following geometrical meaning.
By (\ref{gauge_symmetry_of_omega}), 
the transformation $\bi{z} \mapsto \bi{z}+\epsilon \tilde{\bi{z}}$ ($\tilde{z}_j=[ z_j, C]$) 
gives a co-rotation of $\bi{q}$ and $\bi{p}$ around the axis $\bell$
(note that this rotation is in the space $M_6$, not in the space $M_{\bell}$),
hence, $\bell=\bi{q}\times\bi{p}$ does not change.
The rotation angle can be written as
\[
\theta = \frac{1}{2|\bell|} \tan^{-1} \left( \frac{ ( \bell\times\bi{q})_j}{ q_j |\bell| } \right) 
\]
(we choose the coordinate $q_j\neq0$) and evidently $[\theta, C]=1$.
Let us embed $M_{\bell}$ in the 4-dimensional space $\widetilde{V_{\bell}} = \{ (\bell,\theta) ;\,\bell\in M_{\bell},\,\theta\in {[} 0,2\pi {)} \}$.
For $G(\bell,\theta) \in C^\infty(\widetilde{M_{\bell}})$, we obtain
\[
[G, C ] = \sum_{j=1}^3 \partial_{\ell_j} G [ \ell_j, C] +  \partial_\theta G [ \theta, C] =  \partial_\theta G.
\]
Therefore, the gauge symmetry $[\circ, C]=0$ can be rewritten as $\partial_\theta =0$.
Reversing the view point, for every Hamiltonian $H(\bell,\theta) \in C^\infty(\widetilde{M_{\bell}})$ that has the symmetry $\partial_\theta H=0$,
$C$ is invariant:
\[
\dot{C} = [C, H] = -\partial_\theta H =0.
\]
Therefore, the conjugate variable $\theta$ dictates both the \emph{gauge symmetry} $[\circ , C]=\partial_\theta =0$ of the submanifold $M_\ell \subset M_6$ and the \emph{Casimir symmetry} $\dot{C}= [C, H]=0$ ($\forall H$ such that $\partial_\theta H=0$). 
We can further embed $\widetilde{M_{\bell}}$ in $M_6$ by identifying all canonical variables (see Remark\,\ref{remark:canonical_variables}).

\begin{remark}[Lie-Poisson bracket]
\label{remark:Lie-Poisson}
\normalfont
Given a Lie algebra $\mathfrak{g}$, we can construct a Poisson bracket on the dual space $\mathfrak{g}^*$;
such brackets are called \emph{Lie-Poisson brackets}, because they were known to Lie in the 19th century.
Let $[\;,\;]$ be the Lie bracket of $\mathfrak{g}$, and $\langle \;,\; \rangle$ be the pairing $\mathfrak{g}\times \mathfrak{g}^*\rightarrow\mathbb{K}$ (the field of scalars).
We denote by   $\bmu$ the  vector of $\mathfrak{g}^*$.
For $G(\bmu)\in C^\infty(\mathfrak{g}^*)$, we define its \emph{gradient} $\partial_{\bmu} G \in \mathfrak{g}$ by
\begin{equation}
\delta G = G(\bmu+\epsilon\tilde{\bmu}) - G(\bmu) 
= \epsilon \langle \partial_{\bmu} G, \tilde{\bmu} \rangle + O(\epsilon^2)
\quad (\forall \tilde{\bmu}\in \mathfrak{g}^*).
\label{gradient}
\end{equation}
The dual space $\mathfrak{g}^*$ is made a Poisson manifold by endowing it with
\begin{equation}
\{ G, H \} = \langle  [\partial_{\bmu} G ,\partial_{\bmu} H], \bmu \rangle
= \langle  \partial_{\bmu} G , [\partial_{\bmu} H, \bmu]^* \rangle ,
\label{Lie-Poisson_definition}
\end{equation}
where $[\;,\;]^*:\, \mathfrak{g}\times\mathfrak{g}^*\rightarrow\mathfrak{g}^*$ is the dual representation of $[\;,\;]$.
Because of this construction, $\{ \;,\;\}$ inherits 
bilinearity, anti-symmetry, and the Jacobi's identity from that of $[\;,\; ]$. 
The Leibniz property is explicitly implemented by the derivation $\partial_{\bmu}$,
so $\{\;.\;\}$ is a Poisson bracket.
The forgoing example of  $\mathfrak{so}(3)$, as well as the Vlasov system's Poisson bracket to be formulated in Sec.\,\ref{sec:Vlasov}, are  examples of  Lie-Poisson systems.
\end{remark}

\begin{remark}[complete set of canonical variables]
\label{remark:canonical_variables}
\normalfont
Let us determine  two other canonical variables
(say $\psi$ and $\varphi$)  needed to embed $M_{\bell}$ in $M_6$.
These variables will determine  the gauge freedom of $\bell=\bi{q}\times\bi{p}$;
we demand the canonical relations $[\ell_j, \psi]=[\ell_j, \varphi]=0$
(as well as commutations with $C$ and $\theta$),
which implies $\ad^*_{\psi} \ell_j = \ad^*_{\varphi} \ell_j =0$.
On the surface transverse to $\bell$, $\mathfrak{sl}(2; \mathbb{R})$ has two other actions:
\[
\bi{z} \mapsto \bi{z} + \epsilon (\;0\;,\, \bi{q} ) ,
\quad
\bi{z} \mapsto \bi{z} + \epsilon (\bi{q},\, -\bi{p} ) ,
\]
which correspond to  twist and compression/extension deformations, respectively. 
These transformations can be  generated by the following  pair of conjugate variables:
\[
\psi = \frac{|\bi{q}|^2}{2},
\quad 
\varphi = \frac{\bi{q}\cdot\bi{p}}{|\bi{q}|^2} .
\]
In summary, $(C,\theta,\psi,\varphi)$ span the   complement of the symplectic leaves of the reduced system.
Notice that only $C$ can be represented by the reduced variable $\bell$,
i.e. $C\in C^\infty(M_{\bell})$.
The other parameters $\psi$ and $\varphi$ inflate the phase space to recover $M_6$.
\end{remark}


\section{The ideal fluid system as a sub-algebra of the Vlasov system}
\label{sec:Vlasov}


\subsection{Kinetic Lie-Poisson algebra for the Vlasov system}

Let $\bi{z}=(\bi{x}, \bi{v})=(x^1,\cdots,x^n,v_1,\cdots,v_n)$ 
  be coordinates for a point  of $M=X\times V=\mathbb{T}^n\times\mathbb{R}^n$,
the phase space of a particle,
which is the cotangent bundle $T^* X$ of a configuration space $X$.
For   convenience, we call $X$ the $\bi{x}$-space, and $V$ the $\bi{v}$-space.
 
We call a real-valued function $\psi(\bi{z})\in C^\infty(M)$ an observable, and the   
 space $C^\infty(M)$ is endowed with the Poisson bracket
\begin{equation}
[\psi, \varphi]  = \sum_{j=1}^n (\partial_{x^j} \psi) \, (\partial_{v_j} \varphi) 
- (\partial_{v_j} \psi) \,  (\partial_{x^j} \varphi) ,
\label{kinetic_Lie-bracket}
\end{equation}
where we denote $\mathfrak{g} = C^\infty_{[\;,\;]}(M)$.
The adjoint representation $\ad_h = [ \circ, h]$ of this Lie algebra
describes the Hamiltonian dynamics of a particle, i.e.,  
\[
\dot{\psi} = [ \psi, h],
\]
where $h$ is the particle Hamiltonian. 

The dual space $\mathfrak{g}^*$ is the set of \emph{distribution functions}; 
for an observable $\psi\in \mathfrak{g}$ and a distribution function $f\in\mathfrak{g}^*$, 
\begin{equation}
\langle \psi, f \rangle = \int_M \psi(z) f(z)\, \rmd z
\label{kinetic_inner-product}
\end{equation}
evaluates the mean value of $\psi$ over the distribution function $f$
(see Remarks\,\ref{remark:pure_state} and \ref{remark:adjoint_action}).

The function space $\mathfrak{g}^*$ of distributions will be the Poisson manifold with  the following construction
(corresponding here  to the phase space $M$ of the examples discussed in Sec.\,\ref{sec:examples}).
On the space $\mathfrak{V}=C^\infty(\mathfrak{g}^*)$
(the set of generalized observables defined for distributions=mixed states; see Remark\,\ref{remark:pure_state}),
 the Poisson-Vlasov  Lie-Poisson bracket\cite{Morrison1980,Morrison1982} 
(see also Refs.\;\onlinecite{MW1982,IBB1984,MMW1984,Lainz2019}) is defined as follows:
\begin{equation}
\{ G,H \} = \langle [\partial_f G, \partial_f H], f \rangle ,
\label{kinetic_Lie-Poisson}
\end{equation}
where $\partial_f H \in T^*{\mathfrak{V}} =\mathfrak{g}$ 
is the gradient of $H\in \mathfrak{V}$ (see Remark\,\ref{remark:Lie-Poisson}).
Integrating by parts, we may rewrite (\ref{kinetic_Lie-Poisson}) as
\begin{equation}
\{ G,H \} = \langle \partial_f G, [\partial_f H, f]^* \rangle = \langle \partial_f G, J(f) \partial_f H \rangle,
\label{kinetic_Lie-Poisson-2}
\end{equation}
where $[\;,\;]^*:\,\mathfrak{g}\times\mathfrak{g}^*\rightarrow\mathfrak{g}^*$
evaluates formally as $[a,b]^*=[a,b]$ (see Remark\,\ref{remark:adjoint_action}).
We call  $J(f)\,\circ = [\circ, f]^*$ the Poisson operator.

For $G(f)=\langle \delta(\bi{z}-\bzeta), f(\bi{z})\rangle = f|_{\bi{z}=\bzeta}$, 
Hamilton's equation
$\dot{G} = \{ G, H \}$ evaluates the co-adjoint orbit;
for every point $\bzeta\in M$,
\begin{equation}
 \dot{f} =  [\partial_f H, f]^* ,
\label{Vlasov-0}
\end{equation}
which is the Vlasov equation governing the evolution of the distribution function $f(\bi{z})$ under the action of the \emph{particle Hamiltonian} $h = \partial_f H$.
For example, let 
\[
h(\bi{z}) = \frac{1}{2} |\bi{v}|^2 + \Phi (\bi{x}) ,
\quad H(f) =\frac12 \int_M \left(|\bi{v}|^2 + \Phi(\bi{x})\right) f(\bi{z})\, \rmd z.
\]
where $\Phi$ depends functionally on $f$ via Poisson's equation. 
The first term of $h$ corresponds to the kinetic energy (we set the particle mass to unity), and the second term represents the potential energy (mean field).
Then, (\ref{Vlasov-0}) reads
\[
\dot{f}= \sum_j - \partial_{v_j} h \partial_{x^j} f +\partial_{x^j} h \partial_{v_j} f 
= \sum_j - v_j \partial_{x^j} f + \partial_{x^j} \Phi \partial_{v_j} f.
\] 

\begin{remark}[distribution function]
\label{remark:pure_state}
\normalfont
The dual space $\mathfrak{g}^*$ may be identified as the set of $n$-forms on $M$.
Then, it is better to say that $f\rmd z$  ($\rmd z$ is the phase-space volume element),
or, more generally, a measure on $M$, is the member of the dual space.
However, regarding (\ref{kinetic_inner-product}) as the definition of duality, we may identify the scalar part $f$ as the member of the \emph{dual space} $\mathfrak{g}^*$; 
see Remark\,\ref{remark:adjoint_action} for the identification the dual space as the space of $n$-forms.
The \emph{pure state} $f=\delta(\bi{z}-\bzeta) \in \mathfrak{g}^*$ identifies a point in $M$,
and evaluates $ \langle \psi, f \rangle =\psi(\bzeta)$.
A general $f$ may be regarded as a mixed state.
\end{remark}

\begin{remark}[Hodge duality of $\mathfrak{g}$ and $\mathfrak{g}^*$]
\label{remark:adjoint_action}
\normalfont
A distribution is rigorously a measure on the phase space $M$, 
and is identified as an $n$-form $f^\star := \star f = f \rmd z$,
where $\rmd z$ is the volume form (Lebesgue measure) of $M$, 
$f$ is the scalar part of the distribution,
and $\star$ is the Hodge star operator.
As noted in Remark\,\ref{remark:pure_state}, however, it is often convenient to regard the scalar part $f$ as the distribution function.
Let us denote by $\mathfrak{g}^\star$ the Hodge-dual space of $\mathfrak{g}$,
We may identify $ \mathfrak{g}^*= \star \mathfrak{g}^\star$.
For a scalar (0-form) $\varphi \in \mathfrak{g}$ and an $n$-form $f^\star \in \mathfrak{g}^\star$,
we define $[\varphi, f^\star ]^\star = \star [\varphi, \star f^\star ] = [\varphi, f ]^* \rmd z$.
This $[\;,\; ]^\star:\, \mathfrak{g}\times\mathfrak{g}^\star \rightarrow \mathfrak{g}^\star$ is the original form of the \emph{dual} representation of $[\;,\;]$.
Changing $\star$ to $*$ means that we take the scalar part (Hodge dual) of the distribution
(a \emph{distribution function} is the scalar part of a distribution). 
\end{remark}


\subsection{Reduction to moment variables}
\label{subsec:moment_variables}

As is well known, a ``fluid model'' is derived by taking the $\bi{v}$-space moments of a kinetic model.
Here we review how it works in the framework of Poisson algebras (Hamiltonian mechanics).
For the distribution $f(\bi{z}) \in \mathfrak{g}^*$, we define
 \begin{eqnarray}
\rho(\bi{x},t) &=& \int_V  f (\bi{x},\bi{v},t) \,\rmd^n v,
\label{general-moment-0} 
\\
{P}_j(\bi{x},t) &=& \int_V {v}_j f (\bi{x},\bi{v},t) \,\rmd^n v
\quad (j=1,\cdots,n).
\label{general-moment-1}
\end{eqnarray}
For convenience of notation, we subsume the density $\rho(\bi{x},t)$ in ${P}_\nu(\bi{x},t)$ as the 0-th component.
Using  $v_0 = 1$ as the 0-th component,  we define $n+1$ dimensional co-vector (momentum) $\widehat{\bi{v}}=(v_0, \bi{v})^{\mathrm{T}}$; hence, 
\[
\rho(\bi{x},t) = P_0 (\bi{x},t) = \int_V {v}_0 f (\bi{x},\bi{v},t) \,\rmd^n v.
\]
Therefore, using $\widehat{\bi{P}} =(P_0, \bi{P})^{\mathrm{T}} =(P_0, P_1,\cdots,P_n)^{\mathrm{T}} $ we get the unified representation
\begin{equation}
{P}_\nu (\bi{x},t) = \int_V {v}_\nu f (\bi{x},\bi{v},t) \,\rmd^n v
\quad (\nu=0,\cdots,n).
\label{general-moment-unify}
\end{equation}
We will use a Greek letter (like $\mu$ or $\nu$) for  an index that starts from zero, and Roman letter (like $j$ or $k$) that starts from 1. 
In vector notation, we will put $\widehat{\;}$ when we include a 0th component.

For a functional $G(P_0, P_1,\cdots,P_n)$, the chain rule reads
\begin{equation}
\delta G = \int_M \partial_f G\, \delta f\, \rmd^nv\rmd^nx 
= \int_X \sum_{\nu=0}^n \partial_{{P}_\nu} G\, \delta {P}_\nu \,\rmd^n x .
\label{moment_reduction-0}
\end{equation}
By $\delta {P}_\nu = \int_V {v}_\nu \delta f \,\rmd^nv$ ($\nu=0, 1,\cdots,n$), we obtain
\begin{equation}
\partial_f G =  \sum_{\nu=0}^n  (\partial_{{P}_\nu} G )  {v}_\nu .
\label{moment_reduction}
\end{equation}
For ${g}^\nu := \partial_{{P}_\nu} G$ and ${h}^\nu:= \partial_{{P}_\nu} H$, 
the kinetic Poisson bracket (\ref{kinetic_Lie-bracket}) evaluates as 
\begin{eqnarray*}
[ \partial_f G, \partial_f H] 
&=&
\sum_{j=1}^n \sum_{\nu=0}^n  \partial_{x^j}({g}^\nu {v}_\nu) \, {h}^j - {g}^j \partial_{x^j} ({h}^\nu {v}_\nu) 
\\
&=& [(\bi{h}\cdot\nabla)\bi{g} - (\bi{g}\cdot\nabla)\bi{h}]\cdot \bi{v} + (\bi{h}\cdot\nabla g^0 - \bi{g}\cdot\nabla h^0 ),
\end{eqnarray*}
where $\bi{g}=({g}^1,\cdots,{g}^n)^{\mathrm{T}}$ and $\bi{h}=({h}^1,\cdots,{h}^n)^{\mathrm{T}}$.
Hence, we obtain
\begin{eqnarray}
\{ G, H \} &=& 
\langle [ \partial_f G, \partial_f H] , f \rangle
\nonumber 
\\
&=& \int_X \sum_{\nu=0}^n [(\bi{h}\cdot\nabla){g}^\nu - (\bi{g}\cdot\nabla){h}^\nu]\cdot {P}_\nu \,\rmd^n x
\nonumber
\\
&=& \left( \partial_{\widehat{\bi{P}}} G, J_P(\widehat{\bi{P}}) \partial_{\widehat{\bi{P}}} H \right)
=: \{ G, H \}_P,
\label{moment_reduction-2}
\end{eqnarray}
where the Poisson operator $J_P(\widehat{\bi{P}})$ for  $n=3$ is the Lie-Poisson form  given in  Ref.\;\onlinecite{Morrison-Greene1980}, 
\begin{equation}
 J_P(\widehat{\bi{P}}) = \left( \begin{array}{cc} 
0 & -\nabla \cdot (P_0 \circ ) \\
- P_0 \nabla\;\;\; & -(\nabla\times\bi{P})\times \circ - \bi{P}(\nabla\cdot\circ) - \nabla(\bi{P}\cdot\circ)
\end{array} \right) ,
\label{moment_reduction-3}
\end{equation}
and
\begin{equation}
( \bi{a} , \bi{b} ) = \int \bi{a}(\bi{x})\cdot \bi{b}(\bi{x})\,\rmd^3 x \,.
\label{fluid_inner-product}
\end{equation}


\subsection{Fluid variables}
\label{subsec:fluid_variables}

The bracket in terms of the usual  fluid variables  is derived by changing variables as follows:
\begin{equation}
\widehat{\bi{P}}=(P_0,P_1,\cdots,P_n)\  \leftrightarrow\  \widehat{\bi{U}}=(\rho, U_1,\cdots,U_n),
\label{fluid_variables_transformation}
\end{equation}
where 
\begin{eqnarray}
\rho(\bi{x}) &=& P_0(\bi{x}) = \int f(\bi{x},\bi{v})\,\rmd^3 v,
\label{density} \\
U_j (\bi{x}) &=& \frac{P_j(\bi{x})}{P_0(\bi{x})} =\frac{\int v_j f(\bi{x},\bi{v})\,\rmd^3 v}{\int f(\bi{x},\bi{v})\,\rmd^3 v}
\quad (j=1,2,3).
\label{fluid-velocity}
\end{eqnarray}
The chain rule gives
\begin{eqnarray*}
\delta G &=& \int_X \sum_\nu \partial_{P_\nu} G\,\delta P_\nu \,\rmd^n x
\\
&=& \int_X  \partial_{U_0} G \delta P_0 + \sum_j \partial_{U_\nu} G\left( \frac{\delta P_j}{P_0} - \frac{P_j \delta P_0}{P_0^2} \right) \,\rmd^n x.
\end{eqnarray*}
Hence, we transform
\[
\partial_{P_0} G = \partial_{\rho} G - \frac{1}{\rho} \bi{U}\cdot \partial_{\bi{U}} G,
\quad 
\partial_{\bi{P}} G =  \frac{1}{\rho} \partial_{\bi{U}} G ,
\]
by which we may calculate, for $G(\widehat{\bi{U}})$,
\begin{equation}
\partial_f G =  \partial_\rho G + \sum_{j=1}^n \frac{v_j - U_j}{\rho} \partial_{U_j} G.
\label{fluid_reduction}
\end{equation}
The Poisson bracket (\ref{moment_reduction-2}) transforms into the following \emph{fluid Poisson bracket}:
For $G(\widehat{\bi{U}}), H(\widehat{\bi{U}})$, the Vlasov Lie-Poisson bracket $\{\;,\;\}$ evaluates as
\begin{equation}
\{ G,H \} = \{ G, H \}_F
= (\partial_{\widehat{\bi{U}}} G, J_F (\widehat{\bi{U}}) \partial_{\widehat{\bi{U}}} H ),
\label{bracket-F_derivation}
\end{equation}
where the Poisson operator $J_F (\widehat{\bi{U}}) $ is, when $n=3$, a  form also given in Ref.\;\onlinecite{Morrison-Greene1980}, 
\begin{equation}
 J_F(\widehat{\bi{U}}) = 
\left( \begin{array}{cc} 0 & -\nabla\cdot \\
-\nabla & -\left(\frac{\nabla\times\bi{U}}{\rho}\right) \times
 \end{array} \right)  .
\label{fluid_Poisson}
\end{equation}
We call $\{ G,H \}_F$ the fluid Poisson bracket. 

In fact, the bracket (\ref{fluid_Poisson}) gives the fluid mechanics equations, when we provide it with 
the Hamiltonian composed of  the total  fluid energy; i.e., 
assuming a barotropic internal energy $\mathcal{E}(\rho)$ and an external potential energy $\phi(\bi{x})$, we have 
\begin{equation}
H(\widehat{\bi{U}}) = \int_X \rho \left( \frac{1}{2} |\bi{U}|^2 + \mathcal{E}(\rho) + \phi(\bi{x})  \right)  \,\rmd^3 x.
\label{fluid_Hamiltonain}
\end{equation} 
Then,
\[
\partial_{\widehat{\bi{U}}} H = \left( \begin{array}{c}  \frac{1}{2} |\bi{U}|^2 
+h + \phi \\
\rho \bi{U} \end{array} \right),
\]
where $h=\partial_{\rho} ( \rho \mathcal{E} )$ is the enthalpy.
Then,  Hamilton's equations $\dot{\widehat{\bi{U}}} = J_U(\widehat{\bi{U}}) \partial_{\widehat{\bi{U}}} H $ are the same as the ideal fluid equations,  
\begin{equation}
\left\{ 
\begin{array}{l}
\displaystyle{\partial_t \rho =-  \nabla\cdot(\bi{U} \rho) ,}
\\\;\\
\displaystyle{\partial_t \bi{U} =-(\bi{U}\cdot\nabla)\bi{U} -  \nabla\left( h + \phi  \right),}
\end{array} \right. 
\label{fluid_equation}
\end{equation}
By the thermodynamic definition of pressure, $\mathcal{P}=\rho^2\partial_\rho  \mathcal{E}$, 
we may rewrite $\nabla h=\nabla[ \partial_\rho( \rho  \mathcal{E}) ] = \rho^{-1} \nabla \mathcal{P}$.

In summary, by the \emph{reduction} of the space of kinetic distributions $\mathfrak{g}^* = \{ f(\bi{z}) \}$ to the space of fluid variables $\mathfrak{g}^*_F = \{ \widehat{\bi{U}} = (\rho, \bi{U})^{\mathrm{T}} \}$,
the Vlasov Lie-Poisson algebra $\mathfrak{V}=C^\infty_{\{\;.\;\}}(\mathfrak{g}^*)$ is reduced to a sub-algebra $\mathfrak{V}_F=C^\infty_{\{\;.\;\}_F}(\mathfrak{g}_F^*)$ dictated by the fluid Poisson bracket $\{ G,H \}_F$.
Sometimes it is more convenient to use the equivalent moment variables $\widehat{\bi{P}}$; 
we denote the moment reduction by $\mathfrak{g}^*_P = \{ \widehat{\bi{P}} = (P_0, \bi{P})^{\mathrm{T}} \}$, and the space of moment observables by $\mathfrak{V}_P=C^\infty_{\{\;.\;\}_P}(\mathfrak{g}_P^*)$ .


\subsection{Sub-algebra consisting of linear functions of $v_k$}
\label{subsec:linear_subalgebra}

From (\ref{moment_reduction}), it is evident that $T^* \mathfrak{V}_P$ (or $T^* \mathfrak{V}_F$) consists of only linear functions of $v_k$.
The following Lemma guarantees that the  moment system $\mathfrak{V}_P$ (or, equivalently, the fluid system $ \mathfrak{V}_F$) is a sub-algebra of the Vlasov system $\mathfrak{V}$.

\begin{lemma}[sub-algebra]
\label{lemma:linear_flow}
Let us consider a subset of observables such that
\[
\mathfrak{g}_L = \left\{ \sum_{\nu=0}^n \alpha^\nu(\bi{x}) v_\nu  ;\, \alpha^\nu(\bi{x}) \in C^\infty(X) \right\}.
\]
where $v_1,\cdots,v_n$ are the coordinates of the $\bi{v}$-space, and  $v_0 := 1$.
This $\mathfrak{g}_L$ is a sub-algebra of $\mathfrak{g}$, i.e.
\[
[ \psi,\phi ] \in \mathfrak{g}_L \quad (\forall \psi, \phi \in \mathfrak{g}_L).
\]
\end{lemma}

\noindent
(proof) By direct calculation, we obtain,
for $\psi = \sum_{\nu}\alpha^{\nu}(\bi{x}) v_\nu$ and $\phi = \sum_{\nu}\beta^{\nu}(\bi{x}) v_\nu$,
\[
[\psi,\phi] = \sum_{\nu=0}^n \left( \sum_{j=1}^n \beta^j \partial_{x^j}\alpha^\nu - \alpha_j \partial_{x^j}\beta^\nu \right) v_\nu.
\]
\begin{flushright}
(QED)
\end{flushright}

\bigskip
Notice that $\psi \in \mathfrak{g}_L $ must be a linear function of $v_\nu$, while it may be an arbitrary (smooth) function of $\bi{x}$.  A similar kind of linear reduction was used to describe the Riemann reduction for self-gravitating ellipsoids in Ref.\;\onlinecite{Lebovitz} and for two-dimensional vortices in Ref.\;\onlinecite{pjmFM}. 


\section{Gauge symmetry of the moment (fluid) reduction}
\label{sec:gauge_symmetry}


\subsection{Casimirs and gauge symmetry}
\label{subsec:Casimir}

It is easy to see that the total particle number
\begin{equation}
C_0 =  \int_X \rho(\bi{x})\,\rmd^3x =  \int_M f(\bi{x},\bi{v})\,\rmd^3x\rmd^3v 
\label{Casimir-0}
\end{equation}
is a Casimir of both kinetic and fluid systems (the first expression applies for $\mathfrak{V}_F$ and the second for $\mathfrak{V}$):  because $\partial_\rho C_0=1$ and  $\partial_f C_0=1$,  evidently,  $\{C_0, H\}_F=0$ and  $\{C_0, H\}=0$,  
for every $H \in \mathfrak{V}_F$  and $H \in \mathfrak{V}$, respectively. 


The \emph{helicity}
\begin{eqnarray}
{C} &=& \frac{1}{2} \int \bi{U}\cdot (\nabla\times\bi{U}) \,\rmd^3 x
\nonumber \\
&=& \frac{1}{2} \int \epsilon^{jk\ell} \left( \frac{\int v_j f\, \rmd^3v }{ \int f\, \rmd^3v} \right)
\partial_{x^k} \left( \frac{\int v_\ell f\, \rmd^3v }{ \int f\, \rmd^3v} \right)
\,\rmd^3 x
\label{helicity}
\end{eqnarray}
is a Casimir of the fluid system, but is not a Casimir of  the kinetic system.
In fact, $\partial_\rho C=0$, and
\begin{equation}
\partial_{\bi{U}} {C} =  \nabla\times\bi{U} =: \bOmega ,
\label{helicity_derivative-U}
\end{equation}
(we call $\bOmega$ the \emph{vorticity}),
hence, $\{{C}, H\}_F=0$ for every $H \in \mathfrak{V}_F$.
On the other hand, by (\ref{fluid_reduction}), we obtain
\begin{equation}
\partial_f {C} = \frac{(\bi{v}-\bi{U})\cdot\bi\bOmega}{\rho} ,
\label{helicity_derivative}
\end{equation}
hence, $\{{C}, H\}\neq 0$ for a general $H \in \mathfrak{V}$. 

The constancy of ${C}$ in the fluid system is due to the \emph{gauge symmetry} of the fluid variables, which is implemented through the fluid reduction:

\begin{theorem}[gauge transformation generated by Casimir invariant]
\label{theorem:helicity_transformation}
The co-adjoint action $f \mapsto f+ \epsilon [\partial_f {C}, f ]^* $ generated by the  Casimir (e.g., the helicity)  ${C}$, leaves the 
fluid variables unchanged,
i.e., 
\begin{equation}
\int_V v_\nu  [\partial_f {C}, f ]^* \rmd^n v =0 
\quad (\nu=0,\cdots,n).
\label{gauge-transformation}
\end{equation}
\end{theorem}

\noindent
(proof) As the fluid system $\mathfrak{V}_F$ is a sub-algebra of the Vlasov system $\mathfrak{V}$
(Lemma\,\ref{lemma:linear_flow}),
the Casimir ${C}$, being a constant in $\mathfrak{V}_F$, 
must also be a constant in $\mathfrak{V}$ given that  the Hamiltonian is a function of only the fluid variables $(U_0,\cdots,U_n) =( \rho, \bi{U})$,
or equivalently the moments $(P_0,\cdots,P_n)$.
Therefore,
\[
\dot{{C}} = - \{ H,  {C} \} 
= - \langle \partial_f H , [ \partial_f {C}, f ]^* \rangle 
= -\sum_\nu \langle v_\nu \partial_{P_\nu} H,  [ \partial_f {C}, f ]^* \rangle 
\]
must vanish for all $H(P_0,\cdots,P_n)$.  
Since $\partial_{P_\nu} H$ ($\nu=0,\cdots,n$) only depend on $\bi{x}$, we can write
\[
\langle v_\nu \partial_{P_\nu} H,  [ \partial_f {C}, f ]^* \rangle
= \int_X \partial_{P_\nu} H  \left( \int_V  v_\nu [ \partial_f {C}, f ]^* \rmd^n v \right)\,\rmd^n x.
 \]
Therefore, (\ref{gauge-transformation}) holds.
\begin{flushright}
(QED)
\end{flushright}

\bigskip
Notice that the proof of Theorem\,\ref{theorem:helicity_transformation} only invokes the fact that $C$ is a Casimir (invariant independent of the Hamiltonian) of the sub-algebra $\mathfrak{V}_F$;
we did not use the explicit form of the helicity $C$.
We can also demonstrate (\ref{gauge-transformation}) by direct calculation using the relation (\ref{helicity_derivative}) of the helicity $C$;
let us see how that  works out.
Denoting the perturbation as $\tilde{f}=[ \partial_f {C}, f ]^* $ and putting $\bomega = \bOmega/\rho$, we observe
\begin{eqnarray*}
\tilde{\rho} &=& \int_V \tilde{f}\,\rmd^n v
\\
&=& \int_V \left[ \partial_{\bi{x}} \left(\bomega\cdot {\bi{v}} \right) \cdot \partial_{\bi{v}} f
- \partial_{\bi{x}} \left( \bomega\cdot {\bi{U}} \right) \cdot \partial_{\bi{v}} f
-  {\bomega} \cdot \partial_{\bi{x}} f \right] \,\rmd^n v
\\
&=&
\int_V \left[ - \left( \nabla\cdot {\bomega} \right)  f
-  {\bomega} \cdot \nabla f \right] \,\rmd^n v
\\
&=& - \left( \nabla\cdot {\bomega} \right)  \rho - {\bomega} \cdot \nabla\rho
= - \nabla\cdot \left(\bomega \rho \right) 
= - \nabla\cdot \bOmega  =0.
\end{eqnarray*}
And, for $j=1,2,3$,
\begin{eqnarray*}
\tilde{P}_j &=& \int_V v_j \tilde{f}\,\rmd^n v
\\
&=& \int_V v_j \left[ \partial_{\bi{x}} \left( {\bomega\cdot(\bi{v}-\bi{U})} \right) \cdot \partial_{\bi{v}} f
-   {\bomega} \cdot \partial_{\bi{x}} f \right] \,\rmd^n v
\\
&=&
\int_V- \sum_k \left[ \partial_{v_k} \left( v_j \partial_{x^k} {\bomega\cdot(\bi{v}-\bi{U})} \right)  f
-  {\omega_k} \partial_{x^k} (v_j f) \right] \,\rmd^n v
\\
&=&
\int_V \left[ -\partial_{x^j} \left( {\bomega\cdot(\bi{v}-\bi{U})} \right) 
- v_j \nabla \cdot {\bomega} \right]\,f\,\rmd^n v
-  {\bomega} \cdot \nabla P_j
\\
&=&
- \bi{P}\cdot \partial_{x^j} {\bomega} + \rho\, \partial_{x^j}\left( {\bi{U}\cdot\bomega} \right) 
- P_j \nabla \cdot {\bomega}  -  {\bomega} \cdot \nabla P_j 
\\
&=& \bOmega \cdot (\partial_{x^j} \bi{U} - \nabla U_j )  =0.
\end{eqnarray*}

\begin{remark}[Baroclinic effect]
\label{remark:baroclinic}
\normalfont
The invariance of the fluid variables $(U_0, \cdots, U_3)$ under the gauge-group action
$\ad^*_{\partial_f C} = [\partial_f C, \circ ]^*$
is the reflection of the constancy of the helicity $C$ in the barotropic fluid system $\mathfrak{V}_F$ (a sub-algebra of the Vlasov system $\mathfrak{V}$).
As shown by the forgoing direct calculations, however, the gauge invariance of $(U_0, \cdots, U_3)$ is independent of the fluid model;
even in a baroclinic fluid, in which $C$ is not constant, the action of $\ad^*_{\partial_f C}$ on $f$ does not change $(U_0, \cdots, U_3)$
(whereas it changes the entropy).
To see more precisely how the helicity conservation and the gauge symmetry are related, let us look into the baroclinic effect.
When the internal energy $\mathcal{E}$ depends not only on $\rho$ but also on the specific entropy $\sigma$,
the pressure term in the fluid equation (\ref{fluid_equation})
modifies as $ \rho^{-1}\nabla \mathcal{P} = \nabla h - T \nabla \sigma$ to include the second non-exact term that causes the baroclinic effect
($T=(\partial h/\partial \sigma)_{\mathcal{P}}$ is the temperature).
Then, the helicity obeys
\[
\frac{\rmd}{\rmd t} C = \int_X T \bOmega\cdot\nabla \sigma \,\rmd^3 x.
\]
On the other hand, the gauge transformation of $\sigma=-\int_V f \log f\,\rmd^3v/\rho$ yields (like the foregoing calculations)
\[
\tilde{\sigma} = 
\frac{-\int_V \tilde{f} (\log f +1) \,\rmd^n v}{\rho} - \sigma \frac{\delta \rho}{\rho} 
=
\rho^{-1} \bOmega\cdot\nabla \sigma .
\]
Therefore, $\bOmega\cdot\nabla \sigma =0$ is a generalization of the barotropic condition that makes the helicity $C$ temporally invariant and, at the same time, the specific entropy $\sigma$ gauge invariant (in addition to $(U_0,\cdots, U_3)$).
\end{remark}



\subsection{Casimir of two-dimensional system}
\label{subsec:2D_Casimir}

As noted above, Theorem\,\ref{theorem:helicity_transformation} applies to every Casimir of a sub-algebra.
In a 2-dimensional configuration space ($n=2$), the fluid reduction works out differently, giving rise to a different Casimir.

Embedding $X\subset \mathbb{R}^2$ into $\mathbb{R}^3$, we define the unit normal vector $\bi{e}_\perp$ on $X$.
For a 2-dimensional co-vector $\bi{u}=(u_1, u_2)^{\mathrm{T}}$, we write
$ \bi{e}_\perp\times \bi{u} = (-u_2, u_1)^{\mathrm{T}}$.
In differential geometrical notation, $ \bi{e}_\perp$ is the Hodge * operator that maps a 1-form $\bi{u}=u_1\rmd x^1 + u_2\rmd x^2$ to the $(2-1)$-form $* \bi{u}= u_1\rmd x^2 - u_2\rmd x^1$. 
The \emph{vorticity} is defined by
\[
\bOmega=
\nabla\times\bi{U} =  (\partial_{x^1} U_2-\partial_{x^2} U_1)\bi{e}_\perp =: W\bi{e}_\perp ,
\]
i.e. $\bOmega=(0,0,W)^{\mathrm{T}}$.
Identifying $\bi{e}_\perp = \rmd x^1\wedge\rmd x^2$, $W$ is the exact 2-form $W=\rmd\bi{U}$.
Dividing it by the 2-form $\rho$, we define a scalar $\psi = W/\rho$.

The reduction to the fluid variables $\widehat{\bi{U}}=(\rho, U_1, U_2)^{\mathrm{T}}$ yields
the fluid Poisson operator
\[
J_F = 
\left( \begin{array}{cc} 
0 & -\nabla \cdot  \\
-\nabla\;\;\; & -\psi \bi{e}_\perp \times
\end{array} \right) .
\]
In the 2-dimensional system, 
the helicity $\frac{1}{2}\int_X \bi{U}\cdot\bOmega\,\rmd^3x$
is identically zero,
while its role is played by the following \emph{cross enstrophy}.
For an arbitrary smooth scalar function $g$, we define
\[
C(\widehat{\bi{U}}) = \int_X g(\psi) \rho\,\rmd^2 x .
\]
We easily find that $C(\widehat{\bi{U}})$ is a Casimir, i.e., $J_F \partial_{\widehat{\bi{U}}} C = 0$.
For example, let us take $g(\psi)=\psi^2/2$.  
Then,
\[
\partial_\rho C = -\frac{\psi^2}{2},
\quad
\partial_{\bi{U}} C =  - \nabla_\perp \psi ,
\]
where $\nabla_\perp \psi = \bi{e}_\perp\times  \nabla \psi$, which is identified as the exact $(n-1)$-form $*\rmd \psi$
(here $n=2$).
By (\ref{fluid_reduction}), 
\[
\partial_f C = \partial_\rho C + \frac{\bi{v}-\bi{U}}{\rho}\cdot\partial_{\bi{U}} C
= -\frac{\psi^2}{2} -  \frac{\bi{v}-\bi{U}}{\rho}\cdot \nabla_\perp \psi.
\]
Using this in Theorem\,\ref{theorem:helicity_transformation}, we obtain the following gauge transformation for the 2-dimensional fluid variables:
\[
\int_V v_\nu  [\partial_f {C}, f ]^* \rmd^3 v =0 
\quad (\nu=0,\cdots,2).
\]


\section{Helicity flow and its geometrical meaning}
\label{sec:geometry}


\subsection{Characterization of $\partial_f C$}
\label{subsec:C'}

In Theorem\,\ref{theorem:helicity_transformation}, we have shown that the Casimir (helicity) $C$ generates 
a Hamiltonian flow inducing
the gauge transformation on the distribution function $f$ that preserves the fluid variables $(\rho, \bi{U})$ (or the moments $P_\nu$).
We call the vector $[\circ, \partial_f C]$ the \emph{helicity flow} in the phase space $M$;
its co-adjoint action $[\partial_f C, \circ]^*$ on the distribution function $f$ induces the gauge transformation.
Since the fluid variables are integrals (moments) over the $\bi{v}$-space,
it might be expected that the gauge symmetry pertains to some transformation in the $\bi{v}$-space that does not change the moments.
However, it is not so; the following example shows that the helicity gauge is primarily about the $\bi{x}$-space transformation of $f$:

\begin{example}[linear shear flow]
\normalfont
Suppose that $\rho=1$ and $\bi{U}=x^1 \bi{e}^2$ (a linear shear flow).
Then, $\bomega=\bOmega=\bi{e}^3$, $\partial_f C = v_3$, and hence
\[
\tilde{f}= [\partial_f C , f]^*= - \partial_{x^3} f .
\]
Evidently, the perturbation $\tilde{f}$ does not yield variations in the fluid variables $\rho$ and $\bi{U}$, because they are independent of $x^3$.
\end{example}

This simple example suggests that $C':=\partial_f C$ is basically a momentum-like variable, which is conjugate to the coordinate parallel to $\bomega$. 
When the vector $\bomega$ is not constant, however, $C'$ becomes a \emph{generalized momentum}, mixing coordinates and momenta.
It also contributes a spacial term $\bomega\cdot\bi{U}$ in $C'$; see (\ref{helicity_derivative}).
An interesting analogy of $C'$ and ``canonical momentum'' of magnetized particle will be shown in Theorem\,\ref{theorem:e-2D}. 
Let us study how such a $C'$ generates a transformation in the phase space $M=X\times V$.


\subsection{Helicity flow in the $\bi{v}$-space}
\label{subsec:adjoint_action}

Here we study the geometrical meaning of the helicity flow.
The adjoint operator $\ad_{C'}=[ \circ , C']$, generated by $C'= \partial_f C \in \mathfrak{g}$,
reads as the tangent vector (which we call the helicity flow)
$\sum_{j=1}^n \tilde{x}^j \partial_{x^j} + \tilde{v_j}\partial_{v_j}$ with components
\begin{eqnarray}
\tilde{\bi{x}} &=&  \bomega ,
\label{gauge-transformation-x}
\\
\tilde{\bi{v}} &=& 
-\nabla \big( \bomega \cdot (\bi{v}-\bi{U}) \big) .
\label{gauge-transformation-v}
\end{eqnarray}
In order to elucidate the geometrical meaning of the transformation induced by $[C' , \circ ]$, let us invoke 
differential geometrical notation.
Notice that $\bomega \in TX$ (vector in the $\bi{x}$-space) is defined as $i_{\bomega} \rho = \bOmega$ for the 3-form $\rho$ and 2-form $\bOmega$
(formally we write $\bomega= \bOmega/\rho$ to identify the tangent vector $\bomega$ as the $(2-3)=(-1)$-form).
So, let us call $\bomega$ the \emph{vorticity vector}.
In (\ref{gauge-transformation-v}), $ \bomega \cdot (\bi{v}-\bi{U}) $ is the scalar $i_{\bomega} (\bi{v}-\bi{U})$,
so $\nabla \left( \bomega \cdot (\bi{v}-\bi{U}) \right)$ reads $\rmd (i_{\bomega} (\bi{v}-\bi{U})) \in T^* X$.
By Cartan's formula, we may calculate
\[
\rmd (i_{\bomega} (\bi{v}-\bi{U})) 
= \pounds_{\bomega} (\bi{v}-\bi{U}) - i_{\bomega} \rmd (\bi{v}- \bi{U})
=  \pounds_{\bomega} (\bi{v}-\bi{U}) + i_{\bomega} \rmd \bi{U}\,,
\]
where $ \pounds_{\bomega}$ is the Lie derivative.  For the 2-form $\rmd \bi{U} = \bOmega$, we obtain $i_{\bomega} \rmd \bi{U} = -\bomega\times\bOmega=0$.  
Therefore, we arrive at an illuminating expression 
\begin{equation}
\tilde{\bi{v}} = - \pounds_{\bomega} (\bi{v}-\bi{U}) .
\label{gauge-transformation-v-2}
\end{equation}
Combined with (\ref{gauge-transformation-x}), the adjoint action generated by the helicity is, therefore,
primarily the flow $\bomega $ in $X$ and its reaction $- \pounds_{\bomega} (\bi{v}-\bi{U})$ in $V$.
Notice that  $\bi{v}-\bi{U}$ is the distance of $\bi{v}$ from the \emph{average} $\bi{U}$.

\begin{remark}[gauge transformation for 2-dimensional fluid]
\label{remark:2D_g-enstrophy_symmetry}
\normalfont
Consider the 2-dimensional case (see Sec.\,\ref{subsec:2D_Casimir}).
We define $\bomega = \nabla_\perp\psi/\rho$, which is identified as a vector such that $i_{\bomega} \rho =\rmd \psi$,
i.e., 
\[
\bomega = \frac{1}{\rho} \big[  (\partial_{x^2} \psi) \partial_{x^1} - (\partial_{x^1} \psi) \partial_{x^2} \big].
\]
The adjoint action generated by the cross enstrophy $C$ is $[\partial_f C, \circ] = \sum_j \tilde{x}^j \partial_{x^j} + \tilde{v}_j \partial_{v_j}$ with
\begin{eqnarray}
\tilde{\bi{x}} &=&  \bomega ,
\label{gauge-transformation-x-2D}
\\
\tilde{\bi{v}} &=& 
-\nabla \left[ \bomega \cdot (\bi{v}-\bi{U}) - \frac{\psi^2}{2} \right] 
= -\rmd \left[ i_{\bomega} (\bi{v}-\bi{U}) - \frac{\psi^2}{2} \right].
\label{gauge-transformation-v-2D}
\end{eqnarray}
We find
\[
 i_{\bomega} \rmd \bi{U} =  \psi \rmd  \psi,
\]
hence we obtain
\begin{equation}
\tilde{\bi{v}}  = - \pounds_{\bomega} (\bi{v}-\bi{U}),
\label{gauge-transformation-v-2D}
\end{equation}
which parallels (\ref{gauge-transformation-v-2}) of the 3-dimensional case.
\end{remark}


\subsection{Transformation in $\bi{v}$-space}
\label{subsec:v-space_transformation}

To see how $- \pounds_{\bomega} (\bi{v}-\bi{U})$ works on each fiber $T^*_{\bi{x}}$, we first consider the case when $\rho$ is  constant ($=1$).
Then, $\bomega = \bOmega$ is simply the vector representation of the 2-form $\bOmega$, i.e., 
$i_{\bomega} \vol_x = \bOmega = \rmd \bi{U}$ 
($\vol_x = \rmd x^1\wedge\rmd x^2\wedge\rmd x^3$ is the volume element of $X$).
We   calculate
\begin{equation}
\tilde{\bi{v}}  =- \left( \nabla  \bOmega \right)\cdot (\bi{v}-\bi{U})  + (\bOmega\cdot\nabla) \bi{U}  .
\label{gauge-transformation-v-3}
\end{equation}
Since $\nabla\cdot\bOmega=0$, we have $\mathrm{Tr} \left( \nabla  \bOmega \right) =0$; hence
$\nabla  \bOmega \in \mathfrak{sl}(3,\mathbb{R})$.
Therefore, the first term on the right-hand side of (\ref{gauge-transformation-v-3}) represents a $\bi{v}$-space volume preserving map (epitomized by \emph{rotation}) around the \emph{center} $\bi{U}$. 
The second term $ (\bOmega\cdot\nabla) \bi{U}$ is the displacement of the center $\bi{U}$ induced by the motion $\bomega= \bOmega$ in the $\bi{x}$-space.

Inhomogeneous $\rho$ modifies (\ref{gauge-transformation-v-3}) as
\begin{equation}
\tilde{\bi{v}}  = -\left( \nabla  \bomega \right)\cdot (\bi{v}-\bi{U}) +(\bomega\cdot\nabla) \bi{U} \,,
\label{gauge-transformation-v-4}
\end{equation}
with $\bomega=\bOmega/\rho$.
The role of the second term is the same as the case of $\bomega = \bOmega$.
However, the first term is no longer an $\mathfrak{sl}(3,\mathbb{R})$ action, 
because $\mathrm{Tr} \left( \nabla  \bomega \right) =\nabla\cdot\bomega = \bOmega\cdot \nabla \rho^{-1}$.
We may decompose it as
\begin{eqnarray*}
-\left( \nabla  \bomega \right)\cdot (\bi{v}-\bi{U}) 
&=& -\left[ \frac{1}{\rho} \left( \nabla  \bOmega \right) + \nabla\left( \frac{1}{\rho}  \right)\otimes \bOmega \right]  \cdot (\bi{v}-\bi{U})
\\
&=&
-\frac{1}{\rho} \left( \nabla  \bOmega \right)\cdot (\bi{v}-\bi{U}) -  \bOmega \cdot (\bi{v}-\bi{U}) \nabla\left( \frac{1}{\rho}  \right) ,
\end{eqnarray*}
in which the first term is an $\mathfrak{sl}(3,\mathbb{R})$ action.
The second term adjusts the variation of the density $\rho$ induced by the $\bi{x}$-space motion $\bomega$; 
the $\bi{x}$-space divergence $\sum \partial_{x^j} \tilde{x}^i = \bOmega\cdot\nabla\rho^{-1}$ 
and the $\bi{v}$-space divergence $\sum \partial_{v_j} \tilde{v}_j = -\bOmega\cdot\nabla\rho^{-1}$ cancel each other.


\subsection{Proper volume of $\bi{v}$-space}
\label{subsec:proper_volume}

These observations guide us to the idea of a  \emph{proper metric} (or volume) of the fluid system.
Let us return to the basic relation $i_{\bomega} \rho = \bOmega$.
We may assume that $X$ is not Euclidean, but the metric is deformed by $\rho$ so that
\[
\vol_\rho= \rho\,\rmd x^1\wedge\rmd x^2\wedge\rmd x^3
\]
is the volume form ($\rho$ may be viewed analogous to the  $\sqrt{g}$ of a Riemannian metric). 
Then, we may evaluate the proper-volume divergence as 
\[
\mathrm{div}\,\bomega = (\rmd \, i_{\bomega} \vol_\rho)^* = \rho^{-1} \sum_j \partial_{x^j} ( \rho \omega_j ) 
=  \rho^{-1} \sum_j \partial_{x^j} \Omega_j =0,
\]
implying that the first term $\left( \nabla  \bomega \right)\cdot (\bi{v}-\bi{U}) $ of (\ref{gauge-transformation-v-4}) is a ``$\vol_\rho$  preserving'' map in $V$.
So, the helicity generates a symplectic (thus $M$ space volume preserving) and, at the same time, $\vol_\rho$  preserving group.


\section{Foliation of the kinetic phase space by the helicity flow}
\label{sec:helicity_symmety}


\subsection{Helicity symmetry in the  phase space $M$}

With $C' = \partial_f C $, the co-adjoint action 
\[
\ad^*_{C'} = [C', \circ]^* = -\bomega\cdot\partial_{\bi{x}} + \pounds_{\bomega} (\bi{v}-\bi{U}) \cdot \partial_{\bi{v}}
\]
generates the gauge group that keeps the fluid variables unchanged
(Theorem\,\ref{theorem:helicity_transformation}).
Conversely, if $f$ satisfies
\begin{equation}
\ad^*_{C'} f = [ C', f ]^* =0,
\label{ad^*_C'}
\end{equation}
every Hamiltonian $H(f) \in \mathfrak{V}$ does not change the helicity $C$:
\[
\dot{C}=\{C, H \} = -\{H, C\} =-\langle H', [C',  f]^* \rangle = 0.
\]
We say that $f$ has the \emph{helicity symmetry}, if (\ref{ad^*_C'}) holds.
Then, even if the Hamiltonian $H$ includes non-fluid variables
(for instance, a higher-order moment such as $\int_V g(\bi{v}) f\,\rmd^n v$ with an arbitrary polynomial $g(\bi{v})$)
the system behaves ``fluid-like'' ---  it 
being constrained to lie on  the leaf of $C$ (as well as on that of $C_0$) provided $f$ has the helicity symmetry $[{C'}, f]^* = 0$.
To put it another way, the symmetry breaking $[{C'}, f]^*  \neq 0$ is the necessary condition for the ``kinetic effect'' to manifest as creation/annihilation of the helicity. 
We also note that helicity symmetry $[{C'}, f]^*  = 0$ is NOT a necessary condition for the helicity $C$ to be conserved;
if $H$ only includes fluid variables $P_\nu$, we obtain, by (\ref{gauge-transformation}),
\[
\dot{C}= -\{H, C\}= \sum_\nu \int_X \partial_{P_\nu} H \left( \int_V v_\nu [C', f]^*\rmd^n v \right) \,\rmd^n x =0.
\]
From the practice following Theorem\,\ref{theorem:helicity_transformation}, 
it is evident that higher-moment variables, such as $\int_V g(\bi{v}) f\,\rmd^n v$, are not invariant under the helicity gauge transformation;
hence a non-fluid Hamiltonian including higher moments violates the helicity conservation, if the helicity symmetry is broken.
So, the helicity conservation can be caused by either the helicity symmetry or the fluid reduction.
Seeing the helicity conservation as the litmus test,  the fluid reduction (neglect of higher moments in the Hamiltonian) can be consistent with the kinetic model, if the helicity symmetry holds for the distribution function. 

The aim of this section is to characterize the helicity symmetry in terms of a set of  canonical coordinates for  the phase space $M$.
For a limited class of $\bomega$, we can construct canonical variables $(\alpha,\beta,\gamma, \wp_\alpha, \wp_\beta, \wp_\gamma)$ such that $\wp_\gamma = C'$.
Then, the helicity symmetry means $\ad^*_{C'} f= \partial_\gamma f =0$.
We call such a parameterization of $M$ the \emph{helicity foliation}
(notice the difference from the $C=$ constant leaf in the function space $\mathfrak{g}^*$; {cf.\,Remark\,\ref{remark:FvsC}}).

\subsection{Epi-2D flow}

Suppose that the fluid velocity $\bi{U}$ (a 1-form in the 3-dimensional configuration space) can be parameterized as 
\begin{equation}
\bi{U}= \nabla\varphi + \alpha\nabla\beta.
\label{epi-2D}
\end{equation}
Evidently, such velocity fields constitute a special class of flows, which we have called  \emph{epi-2D}\cite{epi-2D}
(see Remark\,\ref{remark:Clebsch}).
For these flows the helicity is $C=\frac{1}{2}\int \nabla\varphi\cdot\nabla\alpha\times\nabla\beta\,d^3x$, which yields
\begin{eqnarray*}
C' = (\bi{v}-\bi{U})\cdot\bomega &=&
\frac{\bi{v}\cdot\nabla\alpha\times\nabla\beta}{\rho} 
- \frac{\nabla\varphi\cdot\nabla\alpha\times\nabla\beta}{\rho} 
\\
&=&
\frac{\bi{v}\wedge\rmd\alpha\wedge\rmd\beta}{\rho} - \frac{\rmd\varphi\wedge\rmd\alpha\wedge\rmd\beta}{\rho}.
\end{eqnarray*}
Let us denote an element of the Jacobian matrix by $\partial f^i/\partial x^j$,  where $i,j=1,2,3$, 
and the Jacobian determinant by $\partial (f^1,\cdots,f^n)/\partial(x^1,\cdots,x^n)$, $n\leq3$.  For an epi-2D flow, we have

\begin{theorem}[parameterization by epi-2D fluid variables]
\label{theorem:e-2D}
Suppose that, in an open set $W \subset X$,
\[
\frac{\partial(\alpha,\;\beta)}{\partial(x^j, x^k)} \neq 0,
\quad (\exists j, k) .
\]
In a neighborhood $X_{\bi{x}}$ of $\bi{x}\in W$,
there is a scalar $\gamma$ such that 
\begin{equation}
\rmd\alpha\wedge\rmd\beta\wedge\rmd\gamma 
= \frac{\partial(\alpha,\;\beta,\;\gamma)}{\partial(x^1,x^2,x^3)} \,\vol^3_x
= \rho,
\label{Clebsch-1}
\end{equation}
by which we define three independent vectors ($\in TX$)
\[
\bomega_\alpha = \frac{\rmd\beta\wedge\rmd\gamma}{\rho}, \quad
\bomega_\beta = \frac{\rmd\gamma\wedge\rmd\alpha}{\rho}, \quad
\bomega_\gamma = \bomega =\frac{\rmd\alpha\wedge\rmd\beta}{\rho}.
\]
The variables $\alpha,\beta,\gamma$, together with 
\begin{equation}
\wp_\alpha = i_{\bomega_\alpha} \bi{v}- \partial_\alpha\varphi, \quad
\wp_\beta = i_{\bomega_\beta} \bi{v} - \partial_\beta\varphi, \quad
\wp_\gamma = i_{\bomega_\gamma} \bi{v} - \partial_\gamma\varphi, \quad
\label{canonical_momentum}
\end{equation}
constitute canonical coordinates in $W_{\bi{x}}\times V$.
Among them, $\wp_\gamma = \partial_f C$, hence the helicity symmetry is $\partial_\gamma =0$.
\end{theorem}


\bigskip
\noindent
(proof) The third coordinate $\gamma$ can be constructed by solving (\ref{Clebsch-1}) as a hyperbolic PDE.
For instance, assume that $D_1:=\partial(\alpha,\beta)/\partial(x^2,x^3)\neq 0$ in an open set $W_{\bi{x}}$.
Then,  (\ref{Clebsch-1}) can be cast into a first order PDE:
\begin{equation}
\partial_{x^1} \gamma + c_2 \partial_{x^2} \gamma + c_3 \partial_{x^3} \gamma = c_4,
\label{hyperbolicPDE}
\end{equation}
where 
\[
c_2 = \frac{1}{D_1} \frac{\partial(\alpha,\;\beta)}{\partial(x^3,x^1)}, \quad
c_3 = \frac{1}{D_1} \frac{\partial(\alpha,\;\beta)}{\partial(x^1,x^2)}, \quad
c_4 = \frac{1}{D_1} \rho\, .
\]
We can solve (\ref{hyperbolicPDE}) for $\gamma$ by the method of characteristics 
(see examples in Sec.\,\ref{subsec:gamma-example}).

Let us evaluate the kinetic bracket $[\;,\;]$ explicitly.
We may write
\[
[\wp_\gamma, \circ \, ] = \sum_{j} \left( \partial_{x^j} \left( \sum_k \omega_\gamma^k  v_k - \partial_\gamma \varphi \right) \partial_{v_j} 
- \omega_\gamma^j \partial_{x^j} \right).
\]
Evidently, by (\ref{Clebsch-1}), we have 
\[
[\wp_\gamma, \gamma ] = - \frac{\rmd\alpha\wedge\rmd\beta\wedge\rmd\gamma}{\rho}=-1,
\]
as well as
\[
[\wp_\gamma, \alpha ] = - \frac{\rmd\alpha\wedge\rmd\beta\wedge\rmd\alpha}{\rho}=0,
\quad
[\wp_\gamma, \beta ] = - \frac{\rmd\alpha\wedge\rmd\beta\wedge\rmd\beta}{\rho}=0.
\]
For the momentum-like variables, we observe
\begin{eqnarray}
[\wp_\gamma,\wp_\alpha] &=& \sum_j v_j \Big( (\bomega_\alpha\cdot\nabla) \omega_\gamma^j - (\bomega_\gamma \cdot\nabla) \omega_\alpha^j \Big)
\nonumber \\
& & \;\;- \bomega_\alpha\cdot\nabla (\partial_\gamma\varphi)
+ \bomega_\gamma\cdot\nabla (\partial_\alpha\varphi) .
\label{canonical_momentum-1}
\end{eqnarray}
Using a vector calculus formula, let us calculate the Lie derivative $ \pounds_{\bomega_\alpha}\bomega_\gamma$:
\begin{eqnarray*}
& & (\bomega_\alpha\cdot\nabla)\bomega_\gamma -  (\bomega_\gamma\cdot\nabla)\bomega_\alpha
\\
& & \;\;\;\; = \nabla\times(\bomega_\gamma\times\bomega_\alpha) 
+ (\nabla\cdot\bomega_\gamma)\bomega_\alpha - (\nabla\cdot\bomega_\alpha)\bomega_\gamma
\\
& & \;\;\;\; = \nabla\times\left( \frac{1}{\rho} \nabla\beta \right) + 
\nabla \left( \frac{1}{\rho} \right) \times
\left( \frac{ (\nabla\beta\times\nabla\gamma) \times(\nabla\alpha\times\nabla\beta)}{\rho} \right)
\\
& & \;\;\;\; = \nabla \left( \frac{1}{\rho} \right) \times \nabla\beta - \nabla \left( \frac{1}{\rho} \right) \times \nabla\beta =0.
\end{eqnarray*}
Therefore, in (\ref{canonical_momentum-1}), $(\bomega_\alpha\cdot\nabla) \omega_\gamma^j - (\bomega_\gamma \cdot\nabla) \omega_\alpha^j =0$ for every $j$.
On the other hand, we observe
\[
\bomega_\gamma\cdot \nabla \varphi = \frac{\nabla\alpha\times\nabla\beta}{\rho}\cdot
(\partial_\alpha \varphi \nabla\alpha+ \partial_\beta \varphi \nabla\beta + \partial_\gamma \varphi \nabla\gamma) = \partial_\gamma \varphi,
\]
and, similarly, $\bomega_\alpha \cdot\nabla \varphi  = \partial_\alpha \varphi$.
Therefore, the last two terms in (\ref{canonical_momentum-1}) evaluate as 
$\bomega_\alpha\cdot\nabla (\partial_\gamma\varphi)
+ \bomega_\gamma\cdot\nabla (\partial_\alpha\varphi)  
=\partial_\alpha\partial_\gamma \varphi - \partial_\gamma\partial_\alpha\varphi =0$.
In summary, we find  $[\wp_\gamma,\wp_\alpha] =0$.
The permutation $\alpha\rightarrow\beta\rightarrow\gamma\rightarrow\alpha$ yields all other canonical bracket relations.

Finally, notice $\bomega_\gamma = \bomega = \nabla\times\bi{U}/\rho$, and 
\[
 i_{\bomega_\gamma} \bi{U} = \frac{\rmd\alpha\wedge\rmd\beta\wedge\rmd\varphi}{\rho} 
 =  \frac{\partial_\gamma\varphi \, \rmd\alpha\wedge\rmd\beta\wedge\rmd\gamma}{\rho} = \partial_\gamma\varphi .
\]
Hence, $\wp_\gamma = i_{\bomega} (\bi{v}-\bi{U}) = \partial_f C$.
\begin{flushright}
(QED)
\end{flushright}

\bigskip

\begin{remark}[fields vs.\  coordinates]
\label{remark:FvsC}
\normalfont
Some confusion may arise because the variables $(\alpha,\beta,\gamma, \wp_\alpha, \wp_\beta, \wp_\gamma)$ are at once fields (dependent dynamical variables) and coordinates on $M$.  They are fields as is $\bi{U}$ in (\ref{epi-2D}), but in Theorem \ref{theorem:e-2D} they are used as canonical coordinates, which is possible  for any fixed value of the  time variable. 
\end{remark}

Remembering the examples of reductions given in Sec.\,\ref{sec:examples}, we see that
the \emph{helicity symmetry} $\partial_\gamma =0$
 yields the Casimir $\wp_\gamma = C' $ of the reduced ($\gamma$-suppressed) system  ($\subset \mathfrak{g}=C^\infty(M)$).
The helicity $C\in \mathfrak{V}=C^\infty(\mathfrak{g}^*)$ is the integral of
$C'\in \mathfrak{g}=C^\infty(M)$ with respect to the distribution $f$, 
which inherits its invariance from the helicity symmetry in the phase space $M$.
We note that $C$ is a Casimir of the fluid subalgebra $\mathfrak{V}_F$ (Theorem\,\ref{theorem:helicity_transformation}),
whose invariance is due to the wider reduction into the fluid variables,
so the helicity symmetry in $M$ is not a necessary condition for the constancy of $C$ in the fluid system $\mathfrak{V}_F$.
On the contrary, the helicity symmetry guarantees the constancy of $C$ even in the general (non-reduced) Vlasov system $\mathfrak{V}$.

\begin{remark}[Clebsch parameterization and topological charge]
\label{remark:Clebsch}
\normalfont
Representing a 1-form $\bi{U}$ as in (\ref{epi-2D}) is called the \emph{Clebsch parameterization}.
\begin{enumerate}
\item
If $\bi{U}$ is written in the form of (\ref{epi-2D}), the velocity $\bomega= (\nabla\alpha\times\nabla\beta)/\rho$ is \emph{integrable} in the sense that two scalars $\alpha$ and $\beta$ are the integrals of $\bomega$:
\[
\bomega\cdot\nabla\alpha =0, 
\quad
\bomega\cdot\nabla\beta =0.
\]
To represent a general 3-vector, however, we need another pair of parameters $\alpha'$ and $\beta'$ to write\cite{Clebsch}
\begin{equation}
\bi{U}= \nabla\varphi + \alpha\nabla\beta + \alpha'\nabla\beta'.
\label{full-3D}
\end{equation}
Then, $\bomega$ is not necessarily integrable;
the immersion of the orbits of $\ad_{C'}$ may not yield an embedded submanifold in $X$.

\item
The Clebsch parameters $\alpha, \beta, \varphi$ (0-forms), as well as the density $\rho$ (3-form) are dynamical.
In the fluid system $\mathfrak{V}_F$, $\alpha, \beta$, and $\rho$ are Lie-dragged by the fluid velocity $\bi{U}^\dagger\in TX$ 
(the vector counterpart of $\bi{U}\in T^*X$), i.e.
\[
(\partial_t +  \pounds_{\bi{U}^\dagger} ) \alpha =0,\quad  (\partial_t +  \pounds_{\bi{U}^\dagger} ) \beta=0,
\quad 
(\partial_t +  \pounds_{\bi{U}^\dagger} ) \rho =0.
\]
Therefore, the coordinate $\gamma$ is also Lie-dragged, continuously representing the helicity symmetry.
Only $\varphi$ is modified by $(\partial_t + \pounds_{\bi{U}^\dagger} )\varphi =  \frac{1}{2}U^2 -h - \phi$, where  $h$ is the specific enthalpy, and $\phi$ is the potential energy\cite{epi-2D}. 
In the general  dynamics that is generated by $H(f)\in\mathfrak{V}$, however, the Clebsch parameters are no longer dictated only by the fluid variables.

\item
With an arbitrary Lie-dragged scalar $s$
($\gamma$ is a possible choice) and a fluid element $\Omega\subset X$ that moves with the velocity $\bi{U}^\dagger$,
we can define a \emph{charge} 
\[
Q = \int_{\Omega} \rmd \alpha\wedge\rmd\beta\wedge\rmd s,
\]
which is a constant of motion.\cite{epi-2D} 
This $Q$ corresponds to the cross enstrophy.
While the invariance of the helicity $C$ yields only one codimension for the possible dynamics in the function space $\mathfrak{g}^*$,
the invariance of each charge evaluated for arbitrary $\Omega$ poses an infinite number of constraints.

\item

While the invariants $C$ and the $Q$'s belong to $\mathfrak{V}_F$, there are infinitely many codimensions that are separated from $\mathfrak{V}$ in the reduction to the subalgebra  $\mathfrak{V}_F$; see Remark\,\ref{remark:canonical_variables} for analogous examples of such variables in a finite-dimensional system. 
\end{enumerate}

\end{remark}


\subsection{Examples}
\label{subsec:gamma-example}

Consider now some examples for which we can  explicitly display the ``symmetry coordinate'' $\gamma$. 
As usual, we denote the 3-dimensional Cartesian coordinates by $x$, $y$, and $z$.
The essential part of construction is finding  the $\gamma$ that represents the helicity symmetry;
$\varphi$ appears only in the momentum-like variables, so it can be chosen arbitrarily.
We assume $\rho=1$, so that 
\[
\bomega=\nabla\alpha\times\nabla\beta
= \left(  0 , \; \partial_z \alpha,\; -\partial_y \alpha \right)^{\mathrm{T}} , 
\]
and  find that $\alpha$ is the \emph{Gauss potential} of the 2-dimensional vector $(\omega_y, \omega_z)^{\mathrm{T}}$ on the surface $\beta=$ constant.

\begin{example}[elliptic vortex]
\label{example:elliptic}
\normalfont
A simple example is the ellipse:  For positive $a$ and $b$, 
\[
\alpha= a \frac{y^2}{2} + b \frac{ z^2}{2},\;\beta=x \;\Rightarrow \;
\bomega = \left(  0 , \,  bz, \, -ay  \right)^{\mathrm{T}}.
\]
Solving $\bomega\cdot\nabla\gamma=1$, we obtain
\[
\gamma = \frac{1}{\sqrt{a b}} \mathrm{tan}^{-1} \left(\sqrt{\frac{a}{b}} \frac{y}{ z} \right).
\]
Figure\,\ref{fig:eliptic} shows (a) the contours of $\alpha$ and the vector $\bomega$,
and (b) the coordinates $\alpha$ (blue dotted lines) and $\gamma(y,z)$ (black straight lines).
Only when $a=b$ (i.e.\ the circular vortex) are  $\alpha$ and $\gamma$  orthogonal to each other.
The other coordinate $\beta=x$ is orthogonal to both $\alpha$ and $\gamma$.

\begin{figure}
\begin{center}
\;\;\includegraphics[scale=0.5]{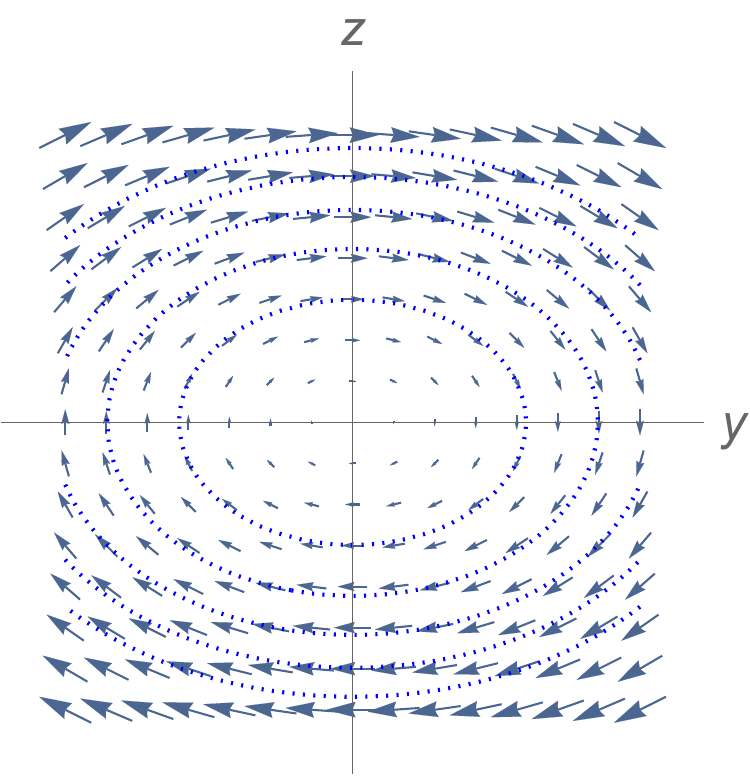}\;\;\;
\includegraphics[scale=0.5]{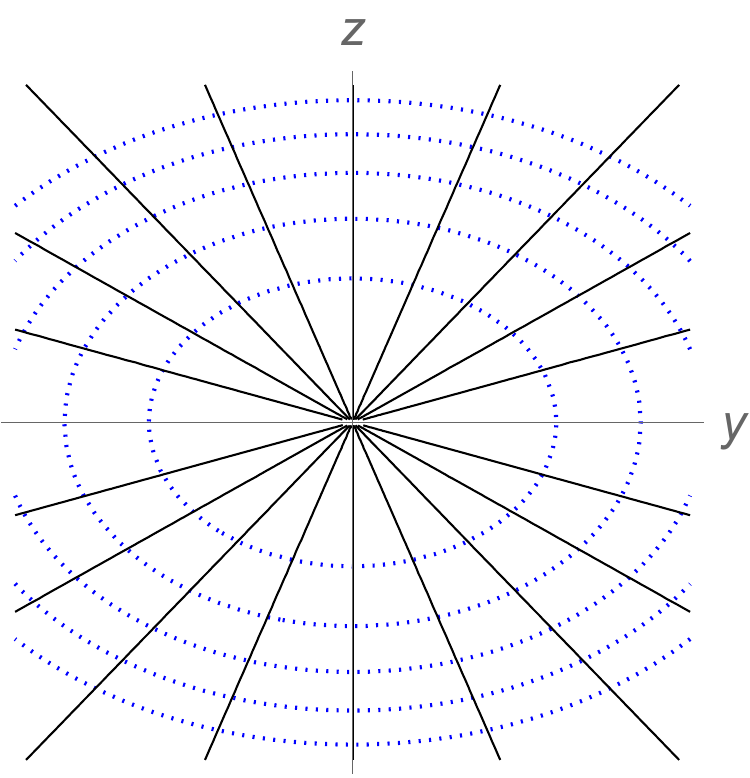}
\\
(a) \;\;\;\;\;\;\;\;\;\;\;\;\;\;\;\;\;\;\;\;\;\;\;\;\;\;\;\;\;\;\;\;\;\;\;\;\;\;\;\;\;\;\;\; (b)
\caption{
(a) Elliptic vortex $\bomega$ and the corresponding Gauss potential $\alpha$ (dotted lines show the level-sets).
(b) The relation between the coordinates $\alpha$ (blue dotted) and $\gamma$ (black).
\label{fig:eliptic}
}
\end{center}
\end{figure}

\end{example}

\begin{example}[hyperbolic vortex]
\label{example:hyperbolic}
\normalfont
As the second example, let us consider the hyperbola (see Fig.\,\ref{fig:hyperbolic} (a)):
\[
\alpha= -\frac{y^2}{2} +  \frac{z^2}{2}, \; \beta =x  \;\Rightarrow \;
\bomega = \left(  0, \, z, \, y \right)^{\mathrm{T}} .
\]
The determining equation for $\gamma$ reads (taking $z$ as the time-like variable)
\begin{equation}
\partial_z \gamma + \frac{z}{y} \, \partial_y \gamma = \frac{1}{ y}.
\label{example-h}
\end{equation}
Upon solving the characteristic equation
\[
\frac{\rmd y}{\rmd z} = \frac{z}{y},
\quad y(0)= y_0,
\]
we obtain $y(z)=\pm\sqrt{z^2+y_0^2}$, or $y_0=\pm\sqrt{y^2-z^2}$.
For an intermediate ``time'' $\zeta$ ($z \geq \zeta \geq 0$), we have
\[
y(\zeta)= \pm\sqrt{y^2-z^2+\zeta^2},
\]
by which we can integrate (\ref{example-h}) as
\[
\gamma = \int_0^z \frac{\rmd \zeta}{ y(\zeta)}.
\]
Because the singularities of the integrand separate different branches of the solution,
we first invoke the indefinite integral: 
\begin{eqnarray}
g(y,z;\zeta) &=& \int \frac{\rmd \zeta}{ y(\zeta)}
\nonumber \\
&=&
\frac{1}{2} \left[ \log \left( 1+ \frac{\zeta}{ y(\zeta)} \right)
 - \log \left( 1- \frac{\zeta}{ y(\zeta)} \right) \right] .
\label{example-h-2}
\end{eqnarray}
Evaluating $g(y,z;\zeta)$ at $\zeta=z$, and setting the ``initial time'' at $\zeta=0$, we obtain
\[
\gamma= \gamma_1 = g(y,z;\zeta)|_0^z = \pm \frac{1}{2} \left[ \log \left( 1+ \frac{z}{\sqrt{y^2}} \right) - \log \left( 1- \frac{z}{\sqrt{y^2}} \right) \right].
\]

\begin{figure}
\begin{center}
\;\;\includegraphics[scale=0.5]{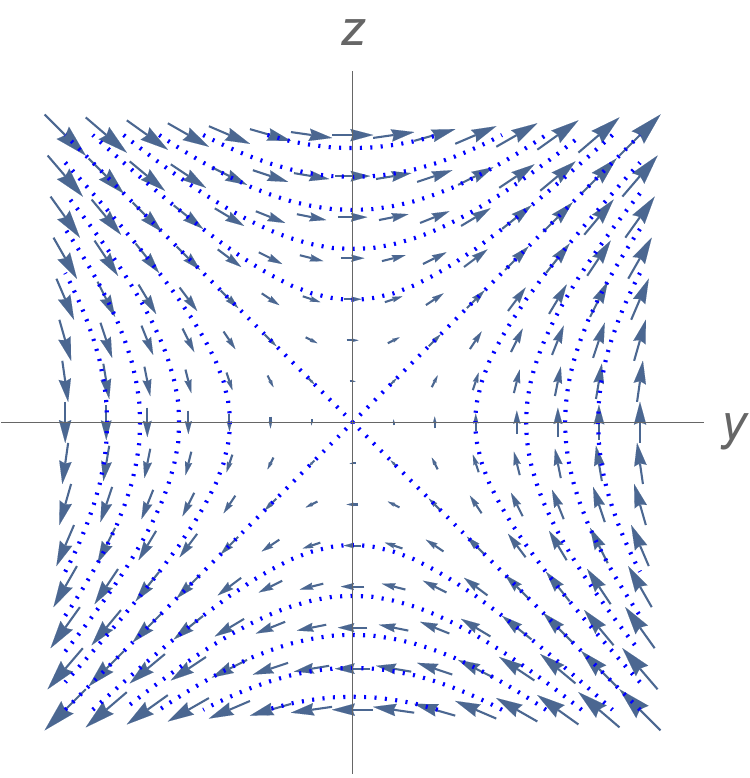}\;\;\;
\includegraphics[scale=0.5]{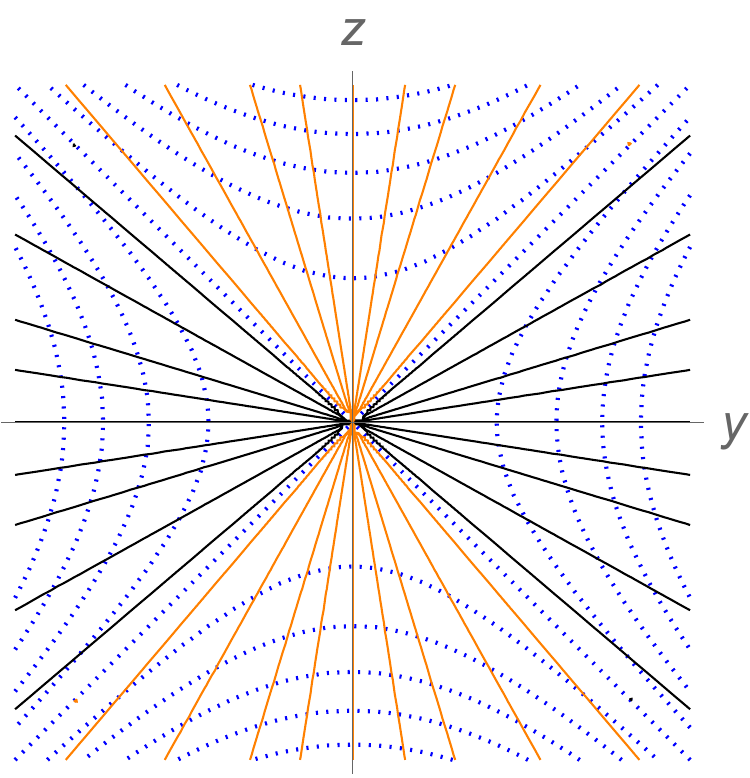}
\\
(a) \;\;\;\;\;\;\;\;\;\;\;\;\;\;\;\;\;\;\;\;\;\;\;\;\;\;\;\;\;\;\;\;\;\;\;\;\;\;\;\;\;\;\;\; (b)
\caption{
(a) Hyperbolic vortex $\bomega$ and the corresponding Gauss potential $\alpha$ (dotted lines show the level-sets).
(b) The relation between the coordinates $\alpha$ (blue, dotted) and $\gamma$ ($\gamma_1$: black, and $\gamma_2$: orange).
\label{fig:hyperbolic}
}
\end{center}
\end{figure}

The characteristic curves that start from $z=0$ do not reach the domain $z^2 > y^2$;
see Fig.\,\ref{fig:hyperbolic} (a).
To construct a solution there, we reverse the roles of $z$ and $y$, and define characteristics for $z|_{y=0} = z_0$.
By the same procedure, we obtain 
\[
\gamma = \gamma_2
= \pm \frac{1}{2} \left[ \log \left( 1+ \frac{y}{\sqrt{z^2}} \right)- \log \left( 1- \frac{y}{\sqrt{z^2}} \right) \right] .
\]
These two functions $\gamma_1$ and $\gamma_2$ define separate local coordinates in the $\bi{x}$-space.
Figure\,\ref{fig:hyperbolic} shows (a) the contours of $\alpha$ and the vector $\bomega$,
and (b) the coordinates $\alpha$ (blue dotted lines) and $\gamma(y,z)$ (black and orange lines).

Here we note that the solution $\gamma$ of the determining equation $\nabla\alpha\times\nabla\beta\cdot\nabla\gamma=1$ is not unique.
Evidently,  the transformation $\gamma \mapsto \gamma + f(\alpha)$  ($f$ an arbitrary $C^1$ function) produces an  infinite set of solutions.
Different choices of such transformations amount to changing the lower-bound of the integral of (\ref{example-h-2}),
because $g(y,z; c)$ ($c$ an arbitrary constant) satisfies $\nabla\alpha\times\nabla\beta\cdot\nabla g(y,z; c)=0$, i.e., $g(y,z; c)=f(\alpha)$.
With the transformation, the boundaries of the coordinate patches move (see Fig.\,\ref{fig:hyperbolic2}).

\begin{figure}
\begin{center}
\;\;\includegraphics[scale=0.5]{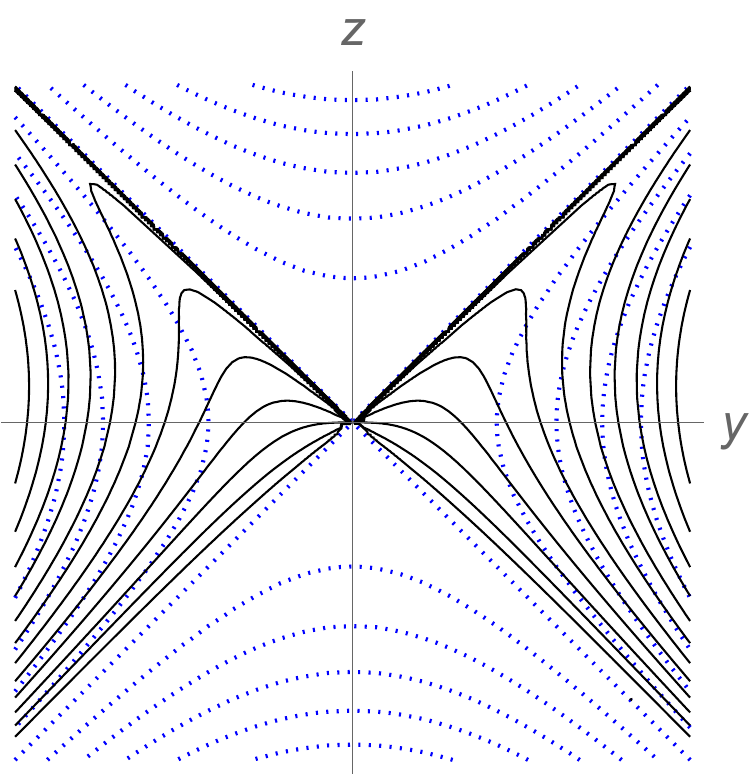}\;\;\;
\includegraphics[scale=0.5]{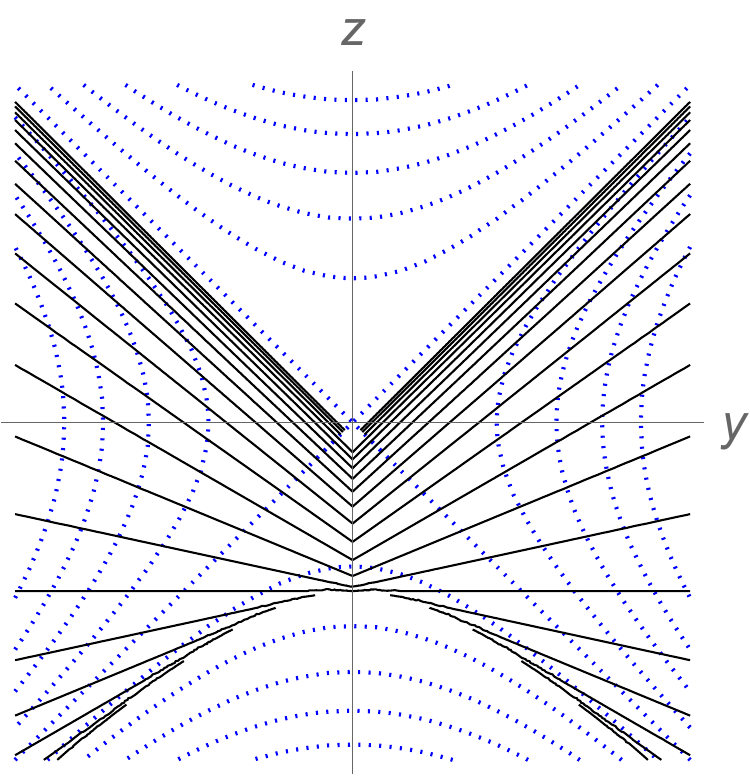}
\\
(a) \;\;\;\;\;\;\;\;\;\;\;\;\;\;\;\;\;\;\;\;\;\;\;\;\;\;\;\;\;\;\;\;\;\;\;\;\;\;\;\;\;\;\;\; (b)
\caption{
(a) Transformed coordinate $g'=g+\alpha$.
(b) Transformation obtained by shifting the lower bound of the integral of (\ref{example-h-2}) to $z=-\pi/2$.
\label{fig:hyperbolic2}
}
\end{center}
\end{figure}

\end{example}

\begin{example}[half an ABC vortex]
\label{example:half-ABC}
\normalfont
A more complicated example is provided by considering  ``half'' of the ABC flow (the example of Ref.\;\onlinecite{epi-2D}; see Fig.\,\ref{fig:half-ABC}):
with three real constants $a,\, b$ and $c$, put
$\varphi=a z \sin x$ and 
\[
\alpha = b \sin y - c \cos z - a z \cos x,\quad
\beta = x
\;\Rightarrow \;
\bomega = \left(  \begin{array}{c} 0 \\ c \sin z -a \cos x \\ -b \cos y  \end{array} \right).
\]
The determining equation for $\gamma$ is,
putting $A=a \cos x$,
\begin{equation}
\partial_z \gamma + \frac{A- c\sin z}{b\cos y}  \partial_y \gamma = -\frac{1}{b\cos y}.
\label{example-2}
\end{equation}
Upon solving the characteristic equation
\[
\frac{\rmd y}{\rmd z} = \frac{A- c\sin z}{b\cos y},
\quad y(0)= y_0,
\]
we obtain, denoting $c'=c/b$ and $A'=A/b$,
\[
y_0 = \sin^{-1} [\sin y + c'(1-\cos z) - A' z]\,.
\]
For an intermediate ``time'' $\zeta$ ($z > \zeta > 0$), we have
\[
y(\zeta) = \sin^{-1} [\sin y + c'(\cos\zeta -\cos z) + A'(\zeta- z)],
\]
by which we integrate (\ref{example-2}) to obtain
\begin{eqnarray}
\gamma &=& -\int_0^z \frac{\rmd \zeta}{b \cos y(\zeta)}
\nonumber\\
&=& -\int_0^z \frac{\rmd \zeta}{\sqrt{b^2-[b\sin y + c(\cos\zeta -\cos z) + a (\zeta- z)\cos x ]^2 }} .
\label{gamma_int}
\end{eqnarray}
Although we cannot evaluate the integral of (\ref{gamma_int}) in terms of elementary functions, it does represent the coordinate $\gamma$ of the helicity symmetry.

\begin{figure}
\begin{center}
\;\;\includegraphics[scale=0.5]{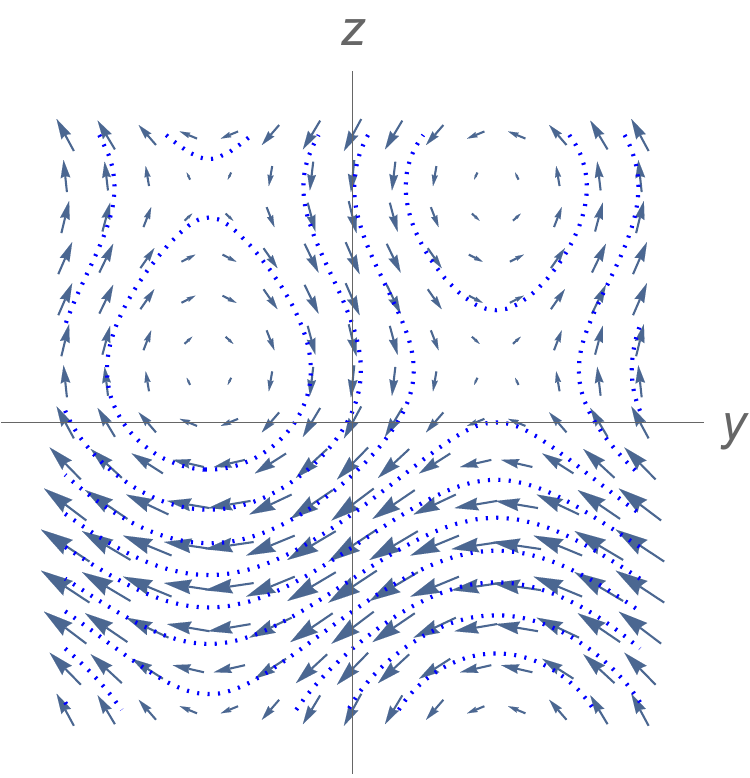}\;\;\;
\includegraphics[scale=0.5]{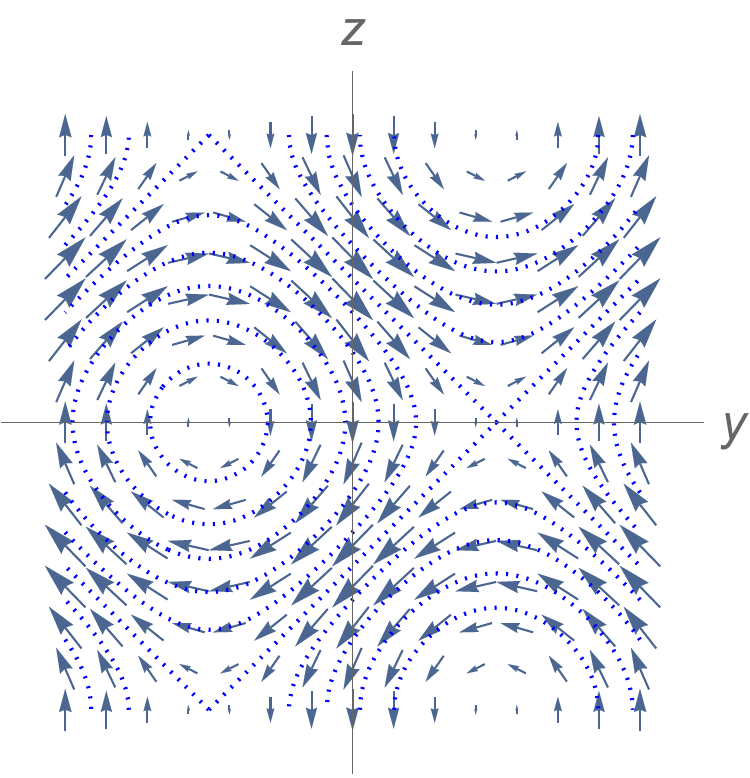}
\\
(a) \;\;\;\;\;\;\;\;\;\;\;\;\;\;\;\;\;\;\;\;\;\;\;\;\;\;\;\;\;\;\;\;\;\;\;\;\;\;\;\;\;\;\;\; (b)
\caption{
``Half ABC''  vortex $\bomega$ and the corresponding Gauss potential $\alpha$ (dotted lines show the level-sets)
on (a) $\beta=x=1$ surface, and (b) $x=\pi/2$.
\label{fig:half-ABC}
}
\end{center}
\end{figure}

For the characteristics curves that do not have an appreciable angle with respect to the $y$-axis, 
we have to choose $y$ as the independent (time-like) variable to rewrite the determining equation (\ref{example-2}) as
\begin{equation}
\partial_y \gamma + \frac{b\cos y}{A- c\sin z} \, \partial_z \gamma = -\frac{1}{A- c\sin z}.
\label{example-2'}
\end{equation}
For $\eta(z) = A z + c \cos z$, we define its inverse function $z(\eta)$ (i.e. $z(\eta(z))=z$);
since $\eta(z)$ is not a monotonic function, $z(\eta)$ needs branch cuts. 
Solving the characteristic equation, we obtain
\[
z(\eta) = z(b\sin\eta - b\sin y + Az + \cos z),
\]
by which we can integrate (\ref{example-2}) as
\[
\gamma = -\int_0^y \frac{\rmd \eta}{A - c \sin z(\eta)}.
\]

Figure\,\ref{fig:ABC-2} shows the coordinates $\alpha$ (blue dotted lines) and $\gamma(y,z)$
(solid lines; different colors indicate different coordinate patches).

\begin{figure}
\begin{center}
\;\;\includegraphics[scale=0.5]{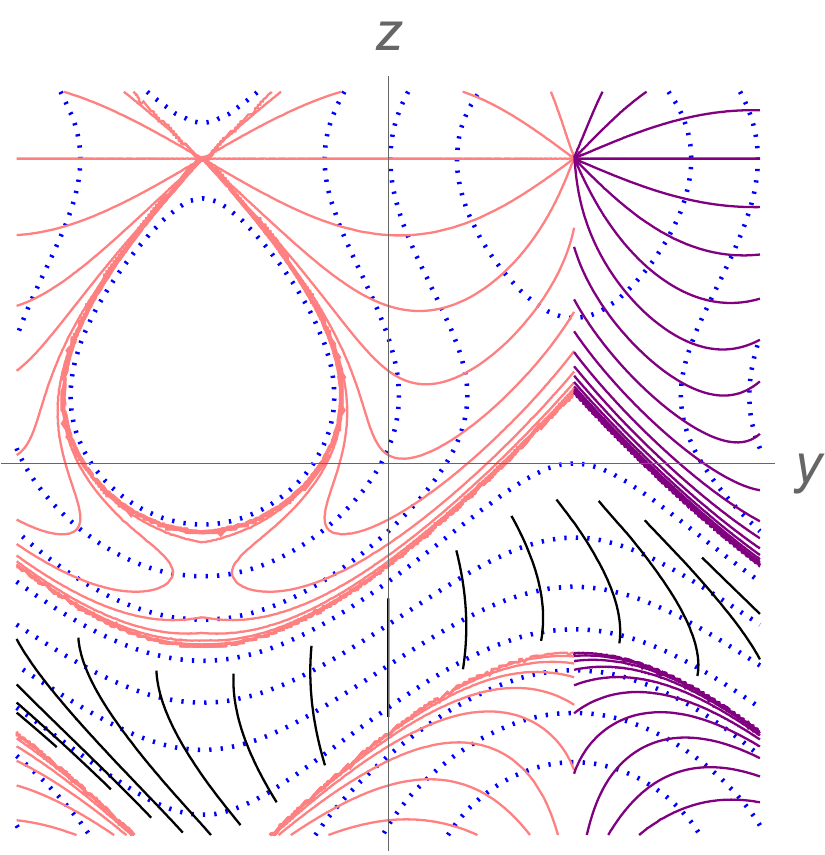}\;\;\;
\includegraphics[scale=0.5]{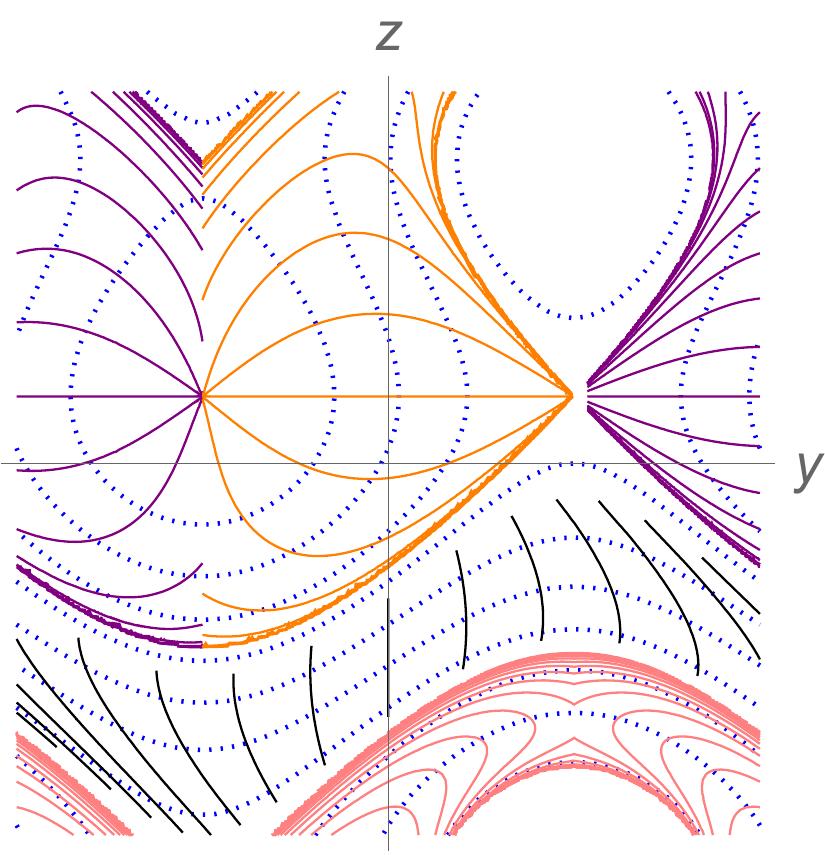}
\\
(a) \;\;\;\;\;\;\;\;\;\;\;\;\;\;\;\;\;\;\;\;\;\;\;\;\;\;\;\;\;\;\;\;\;\;\;\;\;\;\;\;\;\;\;\; (b)
\caption{
The relation between the coordinates $\alpha$ (blue, dotted) and the helicity symmetry coordinate $\gamma$ (different colors indicate different coordinate patches) on the surface $\beta=x=1$.
(a) and (b) show differently patched coordinates.
\label{fig:ABC-2}
}
\end{center}
\end{figure}

\end{example}



\section{Conclusion}

The helicity $C$ is an invariant of the Hamiltonian system governing the fluid variables $(\rho, \bi{U})^{\mathrm{T}}$(or, equivalently, the moments $P_\nu$).
However, it is not an \emph{a priori} invariant in the Vlasov system that dictates the dynamics of the distribution function $f$.
It is the ``reduction'' $f \mapsto P_\nu$ that makes $C$ the Casimir of the fluid system.
Viewed from the Vlasov system, the helicity $C$ represents the gauge symmetry of the fluid variables,
i.e., by $C' = \partial_f C $, the co-adjoint action $\ad^*_{C'} = [C', \circ]^*$ on $f$
generates the gauge group that keeps the fluid variables unchanged
(Theorem\,\ref{theorem:helicity_transformation}).

The topological constraint on vortex lines, which is imposed by the helicity in the fluid system, can be extrapolated to the kinetic Vlasov system as the helicity symmetry. 
If $f$ has the helicity symmetry $\ad^*_{C'} f = [ C', f ]^* =0$,
any Hamiltonian $H(f) \in \mathfrak{V}$ cannot  change the helicity $C$,
i.e. $\dot{C}=\{C, H \} = -\langle H', [C',  f]^* \rangle = 0$.
To put it  another way, the non-symmetry $\ad^*_{C'} f \neq 0$ is the measure of the ``kinetic effect'' that can bring about a change in $C$, unfreezing the topological constraints on vortex lines.
As delineated by Theorem\,\ref{theorem:e-2D},
the helicity symmetry is primarily the homogeneity of $f$ in the direction of the vorticity vector $\bomega$ in $\bi{x}$-space.
If $\bomega$ is not integrable (the case for general 3D flow), the symplectic foliation by the helicity 
(immersion of the orbit of $\ad_{C'}$) may distribute densely in an open set $M_C\subseteq M$ (like the Kronecker foliation).
Then, any inhomogeneity of $f$ in $M_C$ may yield $\ad^*_{C'} f \neq 0$, violating the helicity symmetry.
Even if $\bomega$ is integrable (epi-2D flow), $C'$ is dynamical (see Remark\,\ref{remark:FvsC}),
so it is difficult to maintain $\ad^*_{C'} f =0$ in dynamics generated by an arbitrary Hamiltonian $H$, i.e.,
$[H', C' ]^* \neq 0$ for $H'=\partial_f H$.
We note, however, that the helicity remains constant even when $\ad^*_{C'} f \neq 0$, if the Hamiltonian includes only the fluid variables $P_\nu$, because $\int_V v_\nu \ad^*_{C'} f \,\rmd^3v =0$ for every $f$ (Theorem\,\ref{theorem:helicity_transformation}).

Finally, we note the  remarkable analogy between the Casimir $C$ and the magnetic moment $\mu$ of a magnetized particle (see Sec.\,\ref{subsec:reduction}).
The adiabatic invariance of the action $\mu$ is due to the separation of the microscopic gyration angle $\theta$ from the Hamiltonian.   In the macroscopic model, the homogeneity of the distribution function with respect to $\theta$ justifies the separation of the action $\mu$ and angle $\theta$ variable, resulting is the macroscopic reduced system.   Here, the homogenization of $f$ with respect to the co-adjoint orbit of $\ad^*_{C'}=[C', \circ]^*$ yields $C$ as an adiabatic invariant;
the orbit is in the direction of the vorticity vector $\bomega$ in $X$, accompanied by $- \pounds_{\bomega} (\bi{v}-\bi{U})$ in $V$; see (\ref{gauge-transformation-x}) and (\ref{gauge-transformation-v}).
For an epi-2D flow, we can write $\ad_{C'}=\partial_\gamma$ with conjugate variables $C'=\wp_\gamma$ and $\gamma$ (Theorem\,\ref{theorem:e-2D});
in the analogy of the magnetic moment, $\gamma$ parallels the gyro angle.
As given in (\ref{canonical_momentum}),
the canonical momenta $\wp_j$ ($j=\alpha,\beta,\gamma$) involve the spatial terms $-\partial_j \varphi = - i_{\bomega_j} \bi{U}$,
where $\bi{U}$ resembles the vector potential of the electromagnetic  field.

\acknowledgments
This material is based upon work supported by the National Science Foundation under Grant No.\;1440140, while the authors were in residence at the Mathematical Sciences Research Institute in Berkeley, California, during the semester ``Hamiltonian systems, from topology to applications through analysis'' year of 2018.
The work of ZY was partly supported by JSPS KAKENHI grant number 17H01177,
and that of PJM was supported by the  DOE Office of Fusion Energy Sciences, under DE-FG02-04ER- 54742 and a Forschungspreis from the Alexander von Humboldt Foundation. He warmly acknowledges the hospitality of the Numerical Plasma Physics Division of Max Planck IPP, Garching, Germany,  where a portion of this research was done.

%





\begin{thebibliography}{99}

\bibitem{Moreau}  
J. J. Moreau, 
Constantes d'un \^{i}lot tourbillonnaire en fluide parfait barotrope, 
\textit{C.R.  Acad. Sci. Paris} \textbf{252} (1961)  2810--2812. 

\bibitem{Woltjer}  
L. Woltjer, 
A theorem on force-free magnetic fields, 
\textit{Proc. Nat. Acad.  Sci.}  \textbf{44} (1958) 489--491.
%
\bibitem{Moffatt1969} 
H. K. Moffatt,
The degree of knottedness of tangled vortex lines, 
\textit{J. Fluid Mech.} \textbf{35} (1969) 117--129.
%
\bibitem{Irvine2017} 
M. W. Scheeler, W. M. van Rees, H.  Kedia, D. Kleckner, and W. T. M. Irvine,  
Complete measurement of helicity and its dynamics in vortex tubes, 
\textit{Science} \textbf{357} (2017) 487--491.
%
\bibitem{Irvine2018}   
W. T. M. Irvine, 
Moreau's hydrodynamic helicity and the life of vortex knots and links, 
\textit{Comptes Rendus M\'ecanique} \textbf{346} (2018) 170--174.
%
\bibitem{Biferale2017} 
A. Briard, L. Biferale, and T. Gomez,  
Closure theory for the split energy-helicity cascades in homogeneous isotropic homochiral turbulence,  
\textit{Phys. Rev. Fluids}  \textbf{2} (2017) 102602.
%
\bibitem{Biferale2020} 
P. Clark Di Leoni,  A. Alexakis, L. Biferale and M. Buzzicotti, 
Phase transitions and flux-loop metastable states in rotating turbulence, 
\textit{Phys. Rev. Fluids}  \textbf{5} (2020) 104603.


\bibitem{Morrison1998}
P. J. Morrison,
Hamiltonian description of the ideal fluid,
\textit{Rev.\ Mod.\ Phys.} \textbf{70} (1998)  467--521.

\bibitem{Weinstein1983} 
A. Weinstein,   
The local structure of Poisson manifolds,
\textit{J.  Diff. Geom.}   \textbf{18} (1983)   523--557.

\bibitem{FDR2014}
Z. Yoshida and P. J. Morrison,
A hierarchy of noncanonical Hamiltonian systems: circulation laws in an extended phase space,
\textit{Fluid Dyn. Res.} \textbf{46} (2014)  031412 (21pages).

\bibitem{Marsden}
J. Marsden and A. Weinstein, 
 Reduction of symplectic manifolds with symmetry, 
\textit{Rep. Math. Phys.} \textbf{5}  (1974) 121--130.

\bibitem{Morrison-Greene1980}
P. J. Morrison and J. M. Greene,
Noncanonical {H}amiltonian density formulation of hydrodynamics and Ideal magnetohydrodynamics,
\textit{Phys. Rev. Lett.} \textbf{45} (1980) 790--794.

\bibitem{epi-2D}
Z. Yoshida and P. J. Morrison,
Epi-two-dimensional fluid flow: A new topological paradigm for dimensionality,
\textit{Phys. Rev. Lett.} \textbf{119} (2017)  244501 (5pages).

\bibitem{Northrop} T. G. Northrop, The Adiabatic Motion of Charged Particles (Interscience Publishers, New York, 1963).
%
\bibitem{Henrard} J. Henrard,  The Adiabatic Invariant in Classical Mechanics. In: Jones C.K.R.T., Kirchgraber U., Walther H.O. (eds). Dynamics Reported (Expositions in Dynamical Systems), vol 2. Springer, Berlin, Heidelberg, 1993.

\bibitem{Morrison1980}
P. J. Morrison,
The {M}axwell-{V}lasov equations as a continuous {H}amiltonian system, 
\textit{Phys.  Lett. A} \textbf{80} (1980) 383--386.

\bibitem{Morrison1982}
P. J. Morrison,
Poisson brackets for fluids and plasmas,
\textit{AIP Conf. Proc.} \textbf{88} (1982)  13--46.


\bibitem{MW1982} J. E. Marsden and A. Weinstein, The Hamiltonian structure of the Maxwell-Vlasov equations,
 \textit{Physica D}  \textbf{4} (1982) 394--406.
%
\bibitem{IBB1984} 
I. Bialynicki-Birula and  J. C. Hubbard and L. A. Turski, Gauge-independent canonical formulation of relativistic plasma theory, \textit{Physica A}  \textbf{128} (1984) 509--519.
%
 \bibitem{MMW1984} 
J. E. Marsden, P. J. Morrison, and  A. Weinstein, The Hamiltonian structure of the BBGKY hierarchy equations, \textit{Contemp. Math}  \textbf{28} (1984) 115--124.
%
\bibitem{Lainz2019} 
M. Lainz, C. Sard\'on and  A. Weinstein, Plasma in monopole background does not have a twisted Poisson structure, \textit{Phys. Rev. D}   \textbf{100} (2019)  105016 (5pages).

\bibitem{Lebovitz}
 P. J. Morrison, N. Lebovitz, and J. Biello
The Hamiltonian description of incompressible fluid ellipsoids,
 \textit{ Ann. Physics} {\bf 324} (2009) 1747--1762.
 
\bibitem{pjmFM}
S. Meacham, P. J. Morrison, and G. Flierl, 
Hamiltonian moment reduction for describing vortices in shear,
 \textit{Phys. Fluids} {\bf 9} (1997) 2310--2328.

\bibitem{Clebsch}
Z. Yoshida,
Clebsch parameterization: basic properties and remarks on its applications,
\textit{J. Math. Phys.} {\bf 50} (2009) 113101 (16 pages).

 \end{thebibliography}
\end{document}